\documentclass[12pt]{article}
\usepackage{amsmath,amsfonts,amssymb}
\usepackage{graphicx,color}
\usepackage{bm}
\usepackage{multirow}
\usepackage{cite}
\usepackage{color}
\usepackage[bookmarks, breaklinks, colorlinks,urlcolor=black, citecolor=red, linkcolor=blue]{hyperref}

\usepackage{graphics}
\usepackage{mathrsfs}
\usepackage{slashed}

 \usepackage[normalem]{ulem}

  \long\def\hlpbp#1///{\color{magenta} #1 \color{black}}
 \definecolor{darkgreen}{cmyk}{1,0,1,0.4}

\def\com2#1{\textcolor{red}{\it{#1}}}
\long\def\rpl#1!!#2!!{\color[cmyk]{.3,.5,.9,.1}
  \ensuremath{\rhd} #1 \ensuremath{\lhd}
  \color{blue} #2 \color{black}}

\def\bra{\langle}
\def\ket{\rangle}
\def\del{\partial}
\def\tr{\mathop{\rm Tr}}

\def\Eqn#1{Eq.\ (\ref{#1})}
\def\Eqs#1#2{Eqs.\ (\ref{#1}) and (\ref{#2})}
\def\3Eqs#1#2#3{Eqs.\ (\ref{#1}), (\ref{#2}) and (\ref{#3})}
\def\lag#1{{\mathscr L}_{\rm #1}}

\title{\bf Custodial symmetry, Georgi-Machacek model, and other scalar
  extensions} 

\author{{\bf Anirban Kundu}\thanks{E-mail: akphy@caluniv.ac.in}\,, 
{\bf Poulami Mondal}\thanks{E-mail: poulami.mondal1994@gmail.com}\,, 
{\bf Palash B. Pal}\thanks{E-mail: palashbaran.pal@saha.ac.in} \\ 
Department of Physics, University of Calcutta, \\
92 Acharya Prafulla Chandra Road, Kolkata 700009, India}

\date{November 2021}

\begin{document}

\maketitle

\renewcommand*{\thefootnote}{\fnsymbol{footnote}}


\begin{abstract}

  In an SU(2) gauge theory, if the gauge bosons turn out to be
degenerate after spontaneous symmetry breaking, obviously these mass
terms are invariant under a global SU(2) symmetry that is unbroken.
The pure gauge terms are also invariant under this symmetry.  This
symmetry is called the {\em custodial symmetry} (CS).  In $\rm SU(2)
\times U(1)$ gauge theories, CS implies a mass relation between the
$W$ and the $Z$ bosons.  The Standard Model (SM), as well as various
extensions of it in the scalar sector, possess such a symmetry.  In
this paper, we critically examine the notion of CS and show that there
may be three different classes of CS, depending on the gauge couplings
and self-couplings of the scalars.  Among old models that preserve CS,
we discuss the Two-Higgs Doublet Model and the one doublet plus two
triplet model by Georgi and Machacek.  We show that for two-triplet
extensions, the Georgi-Machacek model is not the most general
possibility with CS.  Rather, we find, as the most general extension,
a new model with more parameters and hence a richer
phenomenology. Some of the consequences of this new model have also
been discussed.

\end{abstract}



\setcounter{footnote}{0}
\renewcommand*{\thefootnote}{\arabic{footnote}}


\section{Introduction}  
The concept of custodial symmetry (CS) is one of the cornerstones of
the Standard Model (SM). CS explains why the $\rho$-parameter, defined
as $M_W^2/M_Z^2\cos^2\theta_W$ where $\theta_W$ is the Weinberg angle,
equals unity at the tree-level in the SM, and also what might be
responsible for a departure from that value when loop corrections to the 
gauge boson self-energy are added. 

To be precise, the Yukawa interaction in the SM breaks the CS and once
the radiative corrections are incorporated, the $\rho$-parameter
shifts from unity, the dominant contribution coming from the top quark
loop and being proportional to $m_t^2$. One defines the parameter
$\rho_0$ to absorb this correction so that in the SM, $\rho_0$ is
unity by definition. The experimental value is close to
unity~\cite{ParticleDataGroup:2020ssz}:
\begin{eqnarray}
\rho_0 = 1.00038 \pm 0.00020\,,
\label{rho-expt}
\end{eqnarray}
so CS must be close to an exact symmetry of nature. This has led to
the construction of possible extensions of the scalar sector of the SM
that respect CS. The Two-Higgs doublet model (2HDM)
\cite{Branco:2011iw} is a well-known example.  It is also
possible to include larger multiplets of scalars in a way that CS is
respected.  The most famous example of this genre is the
Georgi-Machacek (GM) model \cite{Georgi:1985nv}, where a complex
triplet and a real triplet of scalars engender $\rho=1$ at the
tree-level.  Effects and consequences of CS in 2HDM
\cite{Grzadkowski:2010dj, Haber:2010bw}, multi-Higgs doublet models
\cite{Solberg:2018aav}, and GM model \cite{Gunion:1990dt,
  Chiang:2012cn, Blasi:2017xmc, Keeshan:2018ypw} have been discussed
in the literature.  Examples of more CS-conserving models may be found
in Ref.~\cite{Logan:2015xpa}.

When we talk about symmetry, it is necessary to specify its
nature. While the consequence of the CS is to keep $\rho_0=1$, one may
like to know whether this is a symmetry of the entire action, or of
only the gauge-covariant kinetic terms.  For the first case, the
members of every scalar multiplet under the custodial SU(2) should be
degenerate in mass, which is not true for the second case.  It is also
possible to have a model where CS applies to a small part of the
Lagrangian that includes the gauge boson mass terms so that the
relation $\rho_0=1$ is satisfied, but the scalar fields cannot be
grouped into CS multiplets.  These are the cases that we will discuss
and give examples.

We will, in particular, focus on the GM model. After discussing the
nature of CS in the GM model, we will show that this is {\em not} the
most general model with two triplets keeping $\rho_0 = 1$. We
construct the most general model, which we call the extended
Georgi-Machacek (eGM) model, which involves three more parameters in
the scalar sector. 

The eGM model, naturally, has a richer phenomenology than the GM
model. In the second half of the paper, we point out some consequences
of the eGM model. The two major features that may be interesting for
collider physicists are:
\begin{itemize}
\item The scalar multiplets are no longer degenerate and therefore new
decay modes for the scalars open up;
\item The singly charged scalars all develop a nonzero doublet
component and hence can decay to a fermion-antifermion pair.
\end{itemize}

This paper, thus, is organised as follows.  In Section
\ref{sec:custodial} we explain the meaning of the custodial symmetry
(CS) and carefully investigate the link between $\rho_0 = 1$ and CS,
focussing on the SM, 2HDM, and the GM model.  In Section \ref{s:eGM},
we propose and describe the extended GM model, with special attention
to the scalar potential and the scalar mass spectrum. Section
\ref{sec:RGE} enlists the possible theoretical and experimental
constraints on the eGM model.  In Section \ref{sec:results} we provide
a cursory look at the phenomenological aspects of this model,
highlighting the possible signatures that may differentiate this from
the conventional GM model, 
and accommodating hints of several new scalars within the eGM framework.
Section \ref{sec:conclude}  summarizes  and concludes the paper.

\section{Custodial Symmetry}{\label{sec:custodial}}
First, let us define what custodial symmetry (CS) means, and then we
will see how to ensure CS, as well as how the scalar multiplets should
behave under any such CS. We will go through this in a bit detail as
there have been confusing statements about CS in the literature, {\em
e.g.}, whether its role is limited to the gauge boson masses, or
whether it is a symmetry of the scalar sector too.

\subsection{The definition and the ingredients}
The CS should be defined in the context of an SU(2) gauge theory.  It
is well-known that, if such a theory is spontaneously broken by the
vacuum expectation value (VEV) of a complex scalar doublet, the three
gauge bosons acquire equal mass.  We can say that the gauge boson mass
terms possess a global SU(2) symmetry under which the three gauge
bosons form a triplet.  It is trivial to see that the pure gauge terms
are also invariant under this symmetry.  This is called the {\em
  custodial symmetry}.  Inspired by this example, we formulate the
following general definition:
\begin{quote}
  If an SU(2) gauge symmetry undergoes spontaneous symmetry breaking
  (SSB) in a way that the neutral and the charged gauge bosons remain
  mass-degenerate at the tree-level, there is an SU(2) symmetry in the gauge 
  boson mass terms.  The same global symmetry is present in the pure gauge 
  terms even in the broken phase.  This symmetry will be called the 
  {\em custodial symmetry}.

\end{quote}
If we talk about the SM gauge group $\rm SU(2)\times U(1)$, there will
be an extra neutral gauge boson $B$, in addition to the $W_0$ that
comes from SU(2).\footnote{The electric charges will be shown as
  subscripts.}  CS would imply that the $W_0$-$W_0$ term in the
neutral gauge boson mass matrix will be equal to mass-squared value of
the charged gauge bosons.  The other two independent elements of the
matrix are fixed \cite{Palbook} by demanding that the matrix should
have one null eigenvalue, and by calling the angle in the
diagonalizing matrix $\theta_W$.  These conditions imply that the
non-zero mass eigenvalues would satisfy the relation
\begin{eqnarray}
\rho_{\rm tree} = \frac{M_W^2}{M_Z^2\cos^2\theta_W} = 1 \,.
\label{rho=1}
\end{eqnarray}
From now on, we will talk only about tree-level expressions and
relations, unless mentioned otherwise explicitly. Thus, $\rho_0$ and
$\rho_{\rm tree}$ become identical, and we will denote this by
$\rho$. There is a reason for confining ourselves to the tree-level
results; CS for the GM model holds only at the tree-level when
U(1)$_{\rm Y}$ is gauged \cite{Gunion:1990dt}.  We will, therefore,
confine ourselves only to the SU(2) part of the electroweak
interaction, and the consequence of CS will be the mass degeneracy of
the three weak gauge bosons.

Our definition of the CS may be used to test the presence of CS in
extended models as well.  Of course any extension of the SM will not
satisfy the mass relation of \Eqn{rho=1}.  However, among the models
that do, at the tree level at least, there are different kinds,
depending on exactly which parts of the Lagrangian after SSB respect
the CS.  This is the main focus of the present Section.

First, let us write down the condition that ensures $\rho=1$.  Suppose
we have a number of scalar multiplets, to be denoted collectively as
$\Phi$.  The dimensionality of each $\Phi$ is $2T+1$, and the
component that obtains a VEV equal to $v$ has $T_3$-value equal to
$n$.  Then, with an SU(2) theory, one obtains the following masses of
the gauge bosons:\footnote{In this convention, for SM one obtains $M_W
  = gv/\sqrt{2}$, so that $v$ is 173\,GeV.}
\begin{subequations}
  \label{MW}
\begin{eqnarray}
  M_{W_\pm}^2 &=& g^2 \sum c (a_+^2 + a_-^2) v^2 \,,
  \label{MW+} \\
  M_{W_0}^2 &=& 2 g^2 \sum c\, n^2 v^2\,,
  \label{MW0}
\end{eqnarray}
\end{subequations}
where the sum is over all scalar multiplets present in the theory,
with the complexity factor
\begin{eqnarray}
c = \begin{cases} 1 & \mbox{if the multiplet is complex,} \\ \frac12
& \mbox{if the multiplet is real.} 
\end{cases}
\label{c}
\end{eqnarray}
The quantities $a_\pm$ that appear in
\Eqn{MW+} are given by
\begin{eqnarray}
  a_\pm = \sqrt{\frac12\Big[T(T+1) - n(n \pm1)\Big]} \,.
  \label{a+-}
\end{eqnarray}
There are a couple of points associated with \Eqn{MW} that are worth
mentioning here.  First, it is obvious that the two terms in \Eqn{MW+}
come from $|gW_-T_+\bra \Phi \ket|^2$ and $|gW_+T_-\bra \Phi \ket|^2$.
The extra factor of $\frac12$ that appears within the square root sign
of \Eqn{a+-} occurs because of our definition of $T_\pm$ given in
\Eqs{sb.t+-}{sb.tau}, which differ by a factor from what is
encountered in usual textbooks of quantum mechanics.  Thus, adding the
two terms, we can write
\begin{eqnarray}
  M_{W_\pm}^2 = g^2 \sum c\, [T(T+1) - n^2]\, v^2 \,.  
  \label{MWsum}
\end{eqnarray}
While either $a_+$ or $a_-$ vanishes if the VEV occurs in one of the
extreme components of the scalar multiplet, {\em i.e.}, if $n =T$ or
$n = -T$, \Eqn{MWsum} still remains valid, which can be easily
verified by putting $n = \pm T$. Second, the factor $c$ is redundant
in \Eqn{MW0} because real multiplets develop VEV only with $n = 0$ and
therefore do not contribute to the $W_0$-mass.  So, \Eqn{rho=1} will
be satisfied if
\begin{eqnarray}
 \sum c\, [T(T+1) - n^2]\, v^2 = 2 \sum n^2 v^2 \,,
 \label{rho} 
\end{eqnarray}
In the literature, there are examples of many models satisfying
this criterion \cite{Georgi:1985nv, Logan:2015xpa}. 

Even if the relation of \Eqn{rho=1} is realized by a choice of
multiplets through \Eqn{rho}, it will not be respected when higher
order corrections are included.  In particular, it is easy to see that
the fermions will not respect CS.  The fermion mass terms, obtained
after SSB, are bilinear in fields: the left-chiral one being a doublet
and the right-chiral one a singlet.  This cannot be invariant under an
SU(2).  Therefore, in the SM, the CS is realised only on the parts of
the Lagrangian without the fermions. As mentioned before, the Yukawa
terms introduce a deviation of $\rho_{\rm tree}$ from unity, that goes
mostly as $m_t^2$.  In view of this, in the rest of this article we
will completely disregard the fermions, {\em i.e.}, deal only with the
scalars and gauge bosons.  We will see that even the remaining part of
the Lagrangian does not behave in the same way towards CS, and
accordingly different kind of situations might arise regarding masses
of different particles.

We will stick to only the SU(2) part of the SM gauge group
because only this part is relevant for the discussion of CS.  Without
fermions, the Lagrangian density for a set of scalar multiplets
$\{\Phi\}$ can be written as
\begin{eqnarray}
\lag{} = -\textstyle{\frac14} W_{\mu\nu}W^{\mu\nu} + \lag {kin}  -
V(\{\Phi\})\,,
\label{L}
\end{eqnarray}
where $W_{\mu\nu}$ is the SU(2) gauge field tensor and $V(\{\Phi\})$ is
the scalar potential.  The gauge-covariant kinetic term of the scalars
is 
\begin{eqnarray}
  \lag{kin} = \sum  c  \left( D^\mu\Phi \right)^\dag \left(
  D_\mu\Phi   \right) \,,
\end{eqnarray}
where the factor $c$ was defined in \Eqn{c}, the sum is over all
scalar multiplets in the theory, and the gauge-covariant derivative is
defined as
\begin{eqnarray}
D_\mu \Phi = \partial_\mu \Phi + ig \bm t \cdot \bm W_\mu \equiv
\del_\mu \Phi + iG_\mu \Phi \,,
\end{eqnarray}
$\bm t$ being the SU(2) generator matrices in the representation that
$\Phi$ belongs to.  After SSB, the quantum fields $\Phi_0$ with electric charge
equal to zero will need redefinition, for which we will use the
notation 
\begin{eqnarray}
  \Phi_0 = \begin{cases} v + {1 \over \surd2} (\Phi' + i \Phi'') 
    & \mbox{if  the multiplet $\Phi$  is complex,}    \\ 
  v +\Phi'  & \mbox{if  the multiplet $\Phi$  is real},
  \end{cases}
  \label{vev-define}
\end{eqnarray}
Charged fields will not need redefinition.

Clearly, the kinetic term of the gauge bosons is unaffected by SSB,
and will therefore be invariant under a global SU(2) symmetry.  The
$\lag{kin}$ part of the Lagrangian will produce different kinds of
terms after SSB:
\begin{eqnarray}
  \lag{kin} = \lag{mass} + \lag{quad} + \lag{mixed}  +
  \lag{deriv} + \lag{cubic} + \lag{quartic} \,.
  \label{Lkin}
\end{eqnarray}
We now explain the notation, mentioning their relationship with the
CS.
\begin{description}
  \item[$\lag{mass}$:] These are the mass terms of the gauge bosons.
    These terms define CS.  Thus, we are talking only about models
    where these terms respect CS.

  \item[$\lag{quad}$:] These are the kinetic terms of the scalar
    fields, $c(\del^\mu\Phi)^\dagger (\del_\mu\Phi)$.  They have the
    symmetry of the corresponding free field theory.  They need not
    be, and will not be, discussed further.

  \item[$\lag{mixed}$:] These are the terms which contain one power of
    a scalar field and one power of a gauge boson field, {\em i.e.},
    are of the form $(\del^\mu\Phi)^\dagger (iG_\mu \langle \Phi
    \rangle) + \mbox{h.c.}$.  These are the terms which identify the
    unphysical Goldstone fields.  We will see that these terms are
    also CS invariant, with suitably defined CS multiplets.

  \item[$\lag{deriv}$:] These are terms with derivative interactions,
    of the form $(\del^\mu\Phi)^\dagger (iG_\mu \Phi) + \mbox{h.c.}$,
    which contain two powers of scalar fields and one power of gauge
    boson fields.

  \item[$\lag{cubic}$:] These are non-derivative cubic couplings which
    are of the form $(G^\mu\Phi)^\dagger (G_\mu v) + \mbox{h.c.}$.    

  \item[$\lag{quartic}$:] These terms contain two powers of gauge
    fields and two powers of scalar fields, and are of the form
    $(G^\mu\Phi)^\dagger(G_\mu\Phi)$.
    
\end{description}

To summarize, the first kind defines CS and the next two kinds respect
it for sure. Different types of CS will be distinguished by whether
the remaining terms of \Eqn{Lkin}, as well as the scalar potential,
respect CS after SSB.  We list three types, with suitable examples.

\begin{description}
\item [CS-1:] CS is a symmetry of all terms shown in \Eqn{L}.
  Therefore, no correction to $\rho$ is generated even in loops,
  remembering of course that we need to confine ourselves only to the
  Lagrangian of \Eqn{L}, disregarding the fermions.  We will show
  later that this is the case for the SM with only one scalar
  doublet. The canonical GM model is also a CS-1 type model.

\item [CS-2:] In this case, all terms in $\lag{kin}$ respect CS after
  symmetry breaking, but the terms in the scalar potential do not.  We
  will propose a model of this sort in Section~\ref{s:eGM}.  The 2HDM,
  with the most general scalar potential, also falls in this category.

\item [CS-3:] In this case, even $\lag{kin}$ is not fully invariant
  under the CS.  We mentioned before that $\lag{mass}$, $\lag{quad}$
  and $\lag{mixed}$ obey CS.  However, the rest of the terms do not.

\end{description}

\subsection{Some general considerations}
The unphysical Goldstone bosons can be identified by the bilinear
terms containing one scalar field and one gauge boson field, which
arise from $\lag{kin}$ after SSB.  More precisely, after SSB there
will be terms of the form
\begin{eqnarray}
  \lag{mixed} = i M_{W_+} W_+^\mu \del_\mu h_- - i M_{W_-} W_-^\mu \del_\mu
  h_+ - M_{W_0} W_0^\mu \del_\mu h_0 \,,
  \label{Wdh}
\end{eqnarray}
where $h_\pm$ and $h_0$ are the unphysical Goldstone fields.  We will
assume, everywhere in this article, that all VEVs are real.  Then,
\begin{subequations}
  \label{h}
\begin{eqnarray}
  h_+  & = & N \sum c v \left( a_+ \Phi_{(n+1)} - a_- \Phi_{(n-1)}^*
  \right) \,,  
  \label{h+}\\
  h_0 & = & \surd2 N \sum n v \Phi_0'' \,,
  \label{h0} \\ 
  h_-  & = & h_+^* \,.
  \label{h-}
\end{eqnarray}
\end{subequations}
where $\Phi_{(n+1)}$ and $\Phi_{(n-1)}$ denote the components of the
multiplet $\Phi$ just above and below the neutral component (with
$T_3$ eigenvalue equal to $n$), which have electric charges $+1$ and
$-1$ respectively.  If the neutral component of a multiplet happens to
be an extreme one, then one of the terms in \Eqn{h+} would be zero,
and comments made after \Eqn{MWsum} apply for that case.  The
normalization constant $N$ in \Eqn{h} is obvious, and equal in the two
equations owing to the CS relation, \Eqn{rho}.  Note that $h_0$ does
not contain the neutral component of any real multiplet, since these
components have $n=0$ and therefore their VEVs do not break the
neutral generator.

We are considering models with CS in which all the $W$-masses are
equal and the three $W$'s transform as a triplet, with
components\footnote{The minus sign in one of the components
in \Eqn{W3plet}, as well as the factors of $i$ in \Eqn{h3plet}, and
similar such factors in the definitions of various SU(2) multiplets
later in the text, owe their origin to the spherical basis employed.
For details, see Appendix \ref{app:sb}.}
\begin{eqnarray}
\bm W = (-W_+, W_0, W_-) \,.
\label{W3plet}
\end{eqnarray}
If we now define the unphysical Goldstone fields
as part of a triplet
\begin{eqnarray}
\bm h = (ih_+, h_0, ih_-) \,,
\label{h3plet}
\end{eqnarray}
then it is trivial to see that \Eqn{Wdh} can be written as
\begin{eqnarray}
  \lag{mixed} = - M_W \bm W \cdot \del \bm h \,.
  \label{LmixedCS}
\end{eqnarray}
 Here and
henceforth, we employ the following notation for combinations of two CS
multiplets denoted by $\bm a$ and $\bm b$:
\begin{eqnarray}
  \bm a \cdot \bm b &:& \mbox{singlet combination}; \nonumber\\ 
  \bm a \times \bm b &:& \mbox{triplet combination}; \nonumber\\ 
  \bm a \otimes \bm b &:& \mbox{5-plet combination}.
  \label{products}
\end{eqnarray}
The expression in \Eqn{LmixedCS} therefore proves that $\lag{mixed}$
is always invariant under CS, an assertion that was made earlier.
Therefore, among the terms in \Eqn{Lkin}, we need to check only the
status of $\lag{deriv}$, $\lag{cubic}$ and $\lag{quartic}$.

A couple of points can be made irrespective of the scalar multiple
content.  First, the fields associated with the VEVs must be singlets
of CS, since the VEVs are so.  These are the real parts of the neutral
field in any multiplet.  Second, suppose we have a collection of
scalar multiplets in which the highest electric charge in any
component is $Q^{\rm max}$. Then it must be a part of a CS multiplet
of dimension $2Q^{\rm max}+1$.  Numbers of other CS multiplets can be
determined by counting the number of fields of different charges at
our disposal.

\subsection{CS in the Standard Model}\label{s:sm}
Let us first discuss the well-known case of the SM for the groundwork.
In the SM, there is only one Higgs multiplet.  It is a
doublet of SU(2), and will be denoted by $\phi$:
\begin{eqnarray}
  \phi = {\phi_+ \choose \phi_0} \,.
  \label{sm.phi}
\end{eqnarray}
From \Eqn{h}, we see that the unphysical Higgs fields are
\begin{eqnarray}
h_\pm = \phi_\pm, \qquad h_0 = \phi_0'' \,.
\end{eqnarray}
The only left-over field must be a singlet of the CS.  Let us call it
by the suggestive name
\begin{eqnarray}
  S = \phi_0' \,.
\end{eqnarray}
It is now easy to see that, since the VEV is in the direction of
$\phi_0'$ in the field space, there is an SO(3) symmetry in the
remaining directions.  This means that, in the symmetry broken state, 
\begin{eqnarray}
  \phi^\dagger \phi = \phi_+ \phi_- + (v + {1 \over \surd2} \phi_0')^2
  + \frac12 (\phi_0'')^2  = \frac12 \left[ \bm h \cdot \bm h + (S +
    \surd2v)^2  \right] \,.
\end{eqnarray}
This shows that $\phi^\dagger \phi$, and hence the scalar potential,
is CS invariant.

We now look at the terms in \Eqn{Lkin}.  As explained before, only the
last three kinds of terms need to be examined.  It is easy to see that
one can write 
\begin{eqnarray}
    \lag{deriv} &=& \frac12g \Big[ {-} S\bm W \cdot \del \bm h + (\del S)
    \bm W \cdot \bm h - i \bm W \cdot (\bm h \times \del \bm h) \Big]
    \,,  \\
     \lag{cubic} &=& {1 \over 2\surd2} g^2v S \bm W \cdot \bm W \,.
  \label{sm.cubic}
\end{eqnarray} 
The form written here makes it obvious that this part is invariant
under the global CS.  Similarly, the quartic terms can be written as
\begin{eqnarray}
  \lag{quartic} = \frac18 g^2 (\bm W \cdot \bm W) \Big[ \bm h \cdot
    \bm h + S^2 \Big] \,,
  \label{sm.quartic}
\end{eqnarray}
which is also invariant under CS.  Thus, the SM action, even after
SSB, is CS invariant if we neglect the fermions, and it falls under the
CS-1 category.

\subsection{CS in 2HDMs}\label{s:2d}
We now extend our discussion to 2HDM involving two scalar doublets
$\phi_1$ and $\phi_2$, with real VEVs $v_1$ and $v_2$ respectively.
The unphysical fields are now given by
\begin{subequations}
  \label{2d.h}
\begin{eqnarray}
  h_\pm &=& \cos\beta \, \phi_{1\pm} + \sin\beta \, \phi_{2\pm} 
  \,, \\  
  h_0 &=& \cos\beta \, \phi_{10}'' + \sin\beta \, \phi_{20}'' \,,
\end{eqnarray}
\end{subequations}
where $\tan\beta = v_2/v_1$.  As described earlier, these fields
constitute a triplet of a prospective CS.  The orthogonal
combinations\footnote{The neutral component, the CP-odd scalar, is 
conventionally called $A_0$. We use a different notation to display the
multiplet structure in a transparent way.},
\begin{subequations}
  \label{2d.H}
\begin{eqnarray}
  H_\pm &=& {-} \sin\beta \, \phi_{1\pm} + \cos\beta \, \phi_{2\pm}  \,, \\
  H_0 &=& {-} \sin\beta \, \phi_{10}'' + \cos\beta \, \phi_{20}'' \,, 
\end{eqnarray}
\end{subequations}
are then candidates for a triplet of physical Higgs bosons under CS.
In addition, there are two singlets of CS, which are some
combinations of $\phi_{10}'$ and $\phi_{20}'$.  It will be convenient
to define the linear combinations
\begin{eqnarray}
  S_h &=& \cos\beta\, \phi_{10}' + \sin\beta\, \phi_{20}' , \nonumber\\ 
  S_H &=& -\sin\beta\, \phi_{10}' + \cos\beta\, \phi_{20}'\,,
  \label{2d.S}
\end{eqnarray}
even though these may not turn out to be the mass eigenstates.

It can be seen easily that all terms in $\lag{kin}$ are invariant
under the CS.  In particular, we obtain
\begin{eqnarray}
  \lag{deriv} &=&
  \frac12g \bigg[ {-} S_h \bm W \cdot \del \bm h - S_H \bm
    W \cdot \del \bm H + (\del S_h) \bm W \cdot \bm h + (\del S_H) \bm
    W \cdot \bm H \nonumber\\*
    && \qquad -i \bm W \cdot (\bm
    h \times \del \bm h) - i \bm W \cdot (\bm H \times \del \bm H)
    \bigg]
  \,, \\
  \lag{cubic} &=& {1 \over 2\surd2} g^2 \sqrt{v_1^2+v_2^2} \, S_h \bm W \cdot
  \bm W \,, \\
    \lag{quartic} 
    &=& \frac18\, g^2 (\bm W \cdot \bm W) \Big[ \bm h \cdot
          \bm h + \bm H \cdot \bm H + S_h^2 + S_H^2 \Big] \,.      
\end{eqnarray}

In order to check what kind of CS is realized in this
model, we need to look at the scalar potential.  For this, let us
first recall the most general gauge invariant scalar potential of this
model consistent with a discrete symmetry $\phi_1 \to -\phi_1$ in the
quartic sector:
\begin{eqnarray}
\nonumber
V &=& m_{11}^2 \phi_1^\dagger \phi_1+m_{22}^2 \phi_2^\dagger
\phi_2-m_{12}^2(\phi_1^\dagger \phi_2+\phi_2^\dagger
\phi_1) + \frac{\lambda_1}{2}( \phi_1^\dagger
\phi_1)^2 + \frac{\lambda_2}{2}( \phi_2^\dagger \phi_2)^2 \\* 
& & \qquad +\lambda_3 \phi_1^\dagger \phi_1  \phi_2^\dagger \phi_2 +
\lambda_4 \phi_1^\dagger \phi_2  \phi_2^\dagger \phi_1 + 
\frac{\lambda_5}{2}[(\phi_1^\dagger \phi_2)^2+(\phi_2^\dagger
  \phi_1)^2], 
\label{V2HDM}
\end{eqnarray}
This potential has been studied in great detail by many authors
\cite{Branco:2011iw}.  Their results show that the physical states
$H_\pm$ and $H_0$ are degenerate if $\lambda_4=\lambda_5$.  Unless
that condition is fulfilled, we therefore have a CS-2 type symmetry only. 

However, if the condition is fulfilled, the Higgs potential can be
written as
\begin{eqnarray}
\nonumber
V &=& m_{11}^2 \phi_1^\dagger \phi_1+m_{22}^2 \phi_2^\dagger
\phi_2-m_{12}^2(\phi_1^\dagger \phi_2+\phi_2^\dagger
\phi_1) + \frac{\lambda_1}{2}( \phi_1^\dagger
\phi_1)^2 + \frac{\lambda_2}{2}( \phi_2^\dagger \phi_2)^2 \\* 
& & \qquad + \lambda_3 \phi_1^\dagger \phi_1  \phi_2^\dagger \phi_2 +
\frac12 \lambda_4 [\phi_1^\dagger \phi_2 + \phi_2^\dagger 
  \phi_1]^2, 
\label{V2HDMcs}
\end{eqnarray}
Let us now write the different combinations appearing here in terms of
the fields defined in \3Eqs{2d.h}{2d.H}{2d.S} in the Lagrangian after
SSB:
\begin{eqnarray}
  \phi_1^\dagger \phi_1 &=& \frac12 \cos^2\beta \, \bm h \cdot \bm h +
  \frac12 \sin^2\beta \, \bm H \cdot \bm H - \sin\beta\cos\beta \, \bm h
  \cdot \bm H + \frac12 (\phi_{10}' + \surd2 v_1)^2 \nonumber\\ 
  \phi_2^\dagger \phi_2 &=& \frac12 \sin^2\beta \, \bm h \cdot \bm h +
  \frac12 \cos^2\beta \, \bm H \cdot \bm H + \sin\beta\cos\beta \, \bm h
  \cdot \bm H + \frac12 (\phi_{20}' + \surd2 v_2)^2 \nonumber\\
   \phi_1^\dagger \phi_2 + \phi_2^\dagger \phi_1 &=& \frac12 \sin2\beta \,
   (\bm h \cdot \bm h - \bm H \cdot \bm H) +  \cos2\beta \, \bm
   h \cdot \bm H +  (\phi_{10}' + \surd2 v_1) (\phi_{20}' +
   \surd2 v_2) \,. \nonumber\\*
  \label{2d.comb}
\end{eqnarray}
From these forms, it is obvious that the potential of \Eqn{V2HDMcs} is
invariant under CS, which is reflected in the mass degeneracy of
$H_\pm$ and $H_0$.  Therefore, 2HDM, just like SM, is a CS-1 type
model, provided there is a discrete symmetry $\phi_1\to-\phi_1$ in the
potential, and $\lambda_4=\lambda_5$.  If $\lambda_4 \neq \lambda_5$,
or if more terms are admitted in the potential by sacrificing the
discrete symmetry, the model is of CS-2 type.

\subsection{CS in Georgi-Machacek model}\label{ss:CSGM}
In the GM model \cite{Georgi:1985nv}, the scalar sector
consists of a complex doublet $\phi$ as in the SM, a real
triplet $\xi$ and a complex triplet $\chi$ \cite{Keeshan:2018ypw}. The
electric charges of component scalar fields are as follows:
\begin{eqnarray}
\phi = \begin{pmatrix}
\phi_+ \\ \phi_0 \end{pmatrix}, \qquad
\chi = \begin{pmatrix}
\chi_{++} \\ \chi_+ \\ \chi_{0} \end{pmatrix} , \qquad
\xi = \begin{pmatrix}
\xi_{+} \\ \xi_0 \\ -\xi_{-} \end{pmatrix} ,
\label{gm.fields}
\end{eqnarray}
with $\xi_- = \xi_+^*$.  Adopting the notation
\begin{eqnarray}
\bra \phi_0 \ket = v, \qquad 
\bra \chi_0 \ket = u, \qquad 
\bra\xi_0\ket = w, 
\label{gm.vev}
\end{eqnarray}
for the VEVs, one obtains
\begin{eqnarray}
  \rho=\frac{v^2 + 2 \left( u^2 + w^2\right)}{v^2 + 4 u^2}\,.
  \label{gm.rho}
\end{eqnarray}
Thus, $\rho=1$ in the GM model requires $u = w$.

In order to get into a discussion on CS in the scalar potential and
scalar gauge interactions, let us first identify the unphysical Goldstone
fields.  These are 
\begin{subequations}
  \label{gm.h}
\begin{eqnarray}
  h_\pm &=& \cos\beta\, \phi_\pm + {1 \over \surd2} \sin\beta \,
  \left(\chi_\pm + \xi_\pm\right)\,, 
  \label{gm.h+} \\  
  h_0 &=& \cos\beta\, \phi_0'' +  \sin\beta\, \chi_0'' \,,
  \label{gm.h0}
\end{eqnarray}
\end{subequations}
where
\begin{eqnarray}
  \tan\beta = {2u \over v} \,.
  \label{gm.beta}
\end{eqnarray} 

As explained earlier, these three fields must transform like a triplet
of CS.  There will also be a triplet of physical Higgs bosons, given
by~\cite{Georgi:1985nv}
\begin{subequations}
  \label{gm.H}
\begin{eqnarray}
  H_\pm &=&  -\sin\beta \, \phi_\pm + {1 \over \surd2} \cos\beta\, 
  (\chi_\pm + \xi_\pm) \,, \\ 
  H_0 &=&  - \sin\beta\, \phi_0'' + \cos\beta\, \chi_0'' \,.
\end{eqnarray}
\end{subequations}

The doubly charged scalars will be part of a 5-plet
\cite{Georgi:1985nv}.  In our notation, its components are
\begin{subequations}
  \label{gm.F}
\begin{eqnarray}
  F_{\pm\pm} &=& \chi_{\pm\pm} \,, \\ 
  F_\pm &=& {1 \over \surd2} (\chi_\pm - \xi_\pm) \,, \\ 
  F_0 &=& {1 \over \surd3} (\chi_0' - \surd2 \xi_0') \,.
\end{eqnarray}
\end{subequations}
There are also two combinations which are singlets under CS:
\begin{eqnarray}
  S_1 &=& \phi_0',, \qquad S_2 = {1 \over \surd3} (\surd2 \chi_0' +
  \xi_0') \,.
  \label{gm.S}
\end{eqnarray}
There are three multiplets, and yet only two CS singlets.  This is
because there are only two independent VEVs.  Note that the
combination $\chi_0 - \xi_0$ has zero VEV.  Therefore, nothing
prevents the real part of $\chi_0 - \xi_0$ to be part of a non-trivial
CS multiplet.  Indeed, $\mbox{Re} (\chi_0 - \xi_0) = {1 \over
  \surd2}\chi_0' - \xi_0'$ is the combination that appears as $F_0$,
the neutral component of the CS 5-plet.

  We first show explicitly that CS is obeyed by the entire
$\lag{kin}$.  As argued earlier, the first three kinds of terms on the
right side of \Eqn{Lkin} are always CS invariant.  The other terms can
be written as
\begin{eqnarray}
  \lag{deriv} &=& {-} \frac12 g \cos\beta \bigg[ S_1 \bm W \cdot \del
    \bm h - (\del S_1) \bm W \cdot \bm h \bigg] + \frac12 g \sin\beta
  \bigg[ S_1 \bm W \cdot \del \bm H - (\del S_1) \bm W \cdot \bm H
    \bigg] \nonumber\\*
  && - \sqrt{2\over 3} g\sin\beta \bigg[S_2 \bm W \cdot \del \bm h -
    (\del S_2) \bm W \cdot \bm h \bigg]
  - \sqrt{2\over 3} g\cos\beta \bigg[S_2 \bm W \cdot \del \bm H -
    (\del S_2) \bm W \cdot \bm H \bigg] \nonumber\\*
  && - \frac{i}{\sqrt{2}}g \left[ \bm W\cdot \left( \bm h\times \del\bm h + \bm H \times \del\bm H\right)\right]
 +  \sqrt{\frac56} g \sin\beta \, \bm W \cdot \left[\bm F\times \del\bm h - \bm h \times \del\bm F\right]  
  \nonumber\\*
  && + \sqrt{\frac56} g \cos\beta \, \bm W \cdot \left[\bm F\times \del\bm H - \bm H \times \del\bm F\right]  
  + i\sqrt{\frac52}g \bm W \cdot \left( \bm F \times \del \bm F \right)\,, \\
  \lag{cubic} &=& {g^2v \over 2 \surd2} S_1 \bm W \cdot \bm W + {2g^2u \over
    \surd3} S_2 \bm W \cdot \bm W + g^2u \Big( \bm W \otimes \bm W \Big)
  \cdot \bm F \,, \\
  \lag{quartic} &=& g^2 (\bm W \cdot \bm W) \bigg[ \frac18 S_1^2 + \frac13 S_2^2 + 
    \frac13 \bm F \cdot 
    \bm F + \left( \frac18 \cos^2\beta + \frac13 \sin^2\beta \right)
    \bm h \cdot \bm h \nonumber\\*
    && \qquad\qquad + \left( \frac18 \sin^2\beta + \frac13
    \cos^2\beta \right) \bm H \cdot \bm H + \frac5{24} \sin 2\beta \, \bm
    h \cdot \bm H\bigg] \nonumber\\*
   && + g^2 (\bm W \otimes \bm W) \cdot \bigg[ {1
      \over 
    \surd3} S_2 \bm F + {\surd7 \over 4 \surd3} \bm F \otimes
    \bm F + \frac{\surd3}4 \sin\beta\, \bm F \otimes \bm h
    + \frac{\surd3}4 \cos\beta \, \bm F \otimes \bm H \nonumber\\*
    && \qquad\qquad -\frac14 \sin 2\beta\, \bm h \otimes \bm H 
    -\frac14 \sin^2\beta \, \bm h \otimes \bm h
    -\frac14 \cos^2\beta \, \bm H \otimes \bm H \bigg] \,.  
\end{eqnarray}
where the symbol $\bm a \otimes \bm b$ was explained in
\Eqn{products}.  We follow a certain sign convention for the
Clebsch-Gordan coefficients.  With a different convention, the overall
signs might be different the terms where the combinations have been
used.

Let us now turn to the scalar potential.  The most general scalar
potential that is invariant under SU(2)$_{\rm L}\times$U(1)$_{\rm Y}$
has 16 parameters \cite{Keeshan:2018ypw}.  We write it down here, with
some changes in notation which will be helpful in understanding
symmetries of the theory:
\begin{eqnarray}
  V(\phi,\chi,\xi) &=&
  -m_\phi^2 \, \phi^\dag \phi - m_\chi^2\, \chi^\dag \chi -
m_\xi^2\, \xi^\dag \xi \nonumber \\* 
&& +\mu_1 \left(\chi^\dag t_a \chi\right)\, \xi_a + \mu_2 \left(\phi^\dag
\tau_a \phi\right)\, \xi_a  
+ \mu_3 \left[\left(\phi^\top \varepsilon \tau_a \phi\right)\,  \tilde\chi_a  +
  {\rm h.c.}\right] \nonumber\\* 
&&+\lambda_\phi \left(\phi^\dag \phi\right)^2 + \lambda_\xi \left(\xi^\dag
\xi\right)^2 + \lambda_\chi \left(\chi^\dag \chi\right)^2 +  
\tilde{\lambda}_\chi \left\vert \tilde\chi^\dag \chi \right\vert^2
\nonumber\\*  
&& +\lambda_{\phi \chi} \left(\phi^\dag \phi\right) \left(\chi^\dag
\chi\right) + 
\lambda_{\phi \xi} \left(\phi^\dag \phi\right) \left(\xi^\dag \xi\right)  
+ \lambda_{\chi \xi} \left(\chi^\dag \chi\right) \left(\xi^\dag \xi\right)\nonumber\\* 
& & + \kappa_1 \left\vert \xi^\dag \chi  \right\vert^2 + \kappa_2
\left(\phi^\dag \tau_a \phi\right) \left(\chi^\dag t_a \chi\right) 
+ \kappa_3 \left[ \left( \phi^\top \varepsilon \tau_a \phi\right)
  \left(\xi^\dag t_a \tilde\chi \right) + {\rm h.c.} \right]\,, 
\label{gm.V16}
\end{eqnarray}
where
\begin{eqnarray}
\tilde\chi = \begin{pmatrix}
\chi_{0}^* \\
-\chi_- \\ \chi_{--} \end{pmatrix} \,,
\label{gm.chitil}
\end{eqnarray}
which also transforms like a triplet of SU(2), with the U(1) charge
opposite to that of $\chi$.  The 2-dimensional and 3-dimensional
representations of the SU(2) generators are denoted by $\frac12\tau_a$
and $t_a$ respectively.  One may note that these generators are in the
spherical basis, as discussed in Appendix \ref{app:sb},
\3Eqs{sb.t0}{sb.t+-}{sb.tau}, and all of them are not hermitian.
Their hermiticity property is summarized in \Eqn{sb.Tdag}.  A
comparison with earlier notations, {\em e.g.} in
Ref.\ \cite{Keeshan:2018ypw}, has been presented in Appendix
\ref{a:cor}.  Other possible gauge-invariant combinations of four
scalar multiplets can be written as linear combinations of those
appearing in \Eqn{gm.V16}.

This potential will be useful for us for subsequent discussion, but it
does not guarantee equal VEVs of $\chi$ and $\xi$ and therefore the
$W$-mass terms are not degenerate.  In order to have a CS-invariant
potential, Georgi and Machacek used a curtailed potential with 9
parameters, obtained by putting 7 conditions on the parameters of
\Eqn{gm.V16}.  We divide these conditions into two categories, for
reasons to be explained in Section~\ref{s:eGM}.  In the first
category, there are four constraints:
\begin{subequations}
  \label{gm.constr}
\begin{eqnarray}
m_\xi^2 & = & \frac{1}{2}m_\chi^2\,, \\
\mu_2 &=& \surd2 \mu_3 \,, \\
\lambda_\chi & = & 2\lambda_\xi + \frac12 \lambda_{\chi
  \xi}\,, \\ 
\lambda_{\phi\chi} - 2\lambda_{\phi\xi} &=& \kappa_2 +\surd2 \kappa_3\,. 
\label{gm.c4}
\end{eqnarray}
\end{subequations}
The other category has three constraints, which are:
\begin{subequations}
\label{gm.extra}
\begin{eqnarray}
  \kappa_2 + \surd2 \kappa_3 &=& 0 \,,
  \label{gm.c5} \\ 
  \tilde\lambda_\chi = \frac12 \kappa_1 &=& 2 \lambda_\xi - \frac12
  \lambda_{\chi\xi} \,. 
  \label{gm.c6,7}
\end{eqnarray}
\end{subequations}
With these identifications, the GM potential can be written in the
form
\begin{eqnarray}
  V &=& \frac12 m_2^2 \tr (\Phi^\dagger \Phi) + \frac12 m_3^2 \tr
  (X^\dagger X) \nonumber\\*
  && \null - M_1 \tr (\Phi^\dagger \tau_a^\dagger \Phi \tau_b) X_{ab}
  - M_2 \tr (X^\dagger t_a^\dagger X t_b) X_{ab} \nonumber\\*
  && \null + \lambda_1 (\tr \Phi^\dagger \Phi)^2 + \lambda_2 (\tr X^\dagger
  X)^2 + \lambda_3 \tr (X^\dagger X X^\dagger X) \nonumber\\*
  && \null + \lambda_4 (\tr \Phi^\dagger \Phi) \tr (X^\dagger X) -
  \lambda_5 \tr (\Phi^\dagger \tau_a^\dagger \Phi \tau_b) \tr
  (X^\dagger t_a^\dagger X t_b) \,, 
  \label{gm.pot}
\end{eqnarray}
where $\Phi$ and $X$ are two matrices defined as
\begin{eqnarray}
  \Phi = \begin{pmatrix}
    \phi_0^* & \phi_+ \\
    - \phi_- & \phi_0 
  \end{pmatrix} \,, \qquad
  X = \begin{pmatrix}
    \chi_0^* & \xi_+ & \chi_{++} \\
    -\chi_- & \xi_0 &  \chi_+ \\
    \chi_{--} & - \xi_- & \chi_0 
  \end{pmatrix} \,.
  \label{gm.PhiX}
\end{eqnarray}
The parameters of \Eqn{gm.pot} can easily be identified in terms of
the parameters of \Eqn{gm.V16} subject to the conditions of
\Eqs{gm.constr} {gm.extra}.  Such correspondences have been shown in
Appendix \ref{a:cor}.

The question of CS then boils down to checking the CS invariance of
the combinations of fields that appear in \Eqn{gm.pot}.  For example,
by inverting \3Eqs{gm.h}{gm.H}{gm.F}, it is straightforward to see
that
\begin{eqnarray}
  \phi^\dagger \phi = \frac12 \cos^2\beta\, \bm h \cdot \bm h +
  \frac12 \sin^2\beta\, \bm H \cdot \bm H - \sin\beta \cos\beta\, \bm h
  \cdot \bm H + \frac12 (S_1 + \surd2 v)^2 \,,
\end{eqnarray}
which is CS invariant.  As for the other bilinear terms, we find
\begin{subequations}
\begin{eqnarray}
  \chi^\dagger \chi &=& F_{++} F_{--} + 
  \frac12 \bigg| 
  \sin\beta\, h_+ + \cos\beta\, H_+ + F_+
  \bigg|^2 \nonumber\\*
  && \qquad + \frac12 \bigg( 
  \sin\beta\, h_0 +  \cos\beta\, H_0\bigg)^2 
  + \frac12 \bigg( \sqrt{1\over3} F_0 + \sqrt{2\over3} S_2 +
  \surd2 u \bigg)^2 \,, \\
  \xi^\dagger \xi &=& \bigg| \sin\beta\, h_+ 
   + \cos\beta\, H_+ 
   - F_+ \bigg|^2 + \bigg( {-} \sqrt{2\over3}
  F_0 +  \sqrt{1\over3} S_2 + u \bigg)^2 \,.
\end{eqnarray}
\end{subequations}
None of these expressions is CS invariant.  However, the combination
that appears in \Eqn{gm.pot} is
\begin{eqnarray}
  \chi^\dagger \chi + \frac12  \xi^\dagger \xi = \frac12 \left[ \bm F
    \cdot \bm F +  \sin^2\beta\, \bm h \cdot \bm h 
  +  \cos^2\beta\, \bm H 
  \cdot \bm H + \sin 2\beta\, \bm h \cdot \bm H
  + S_2^2\right]\,, 
\end{eqnarray}
leaving out constant terms which are irrelevant and terms linear in
the VEV, since they vanish through the potential minimization
condition.  This is obviously CS invariant.  Similarly CS invariance
can be checked for all other terms in the potential.   The GM
model is therefore of type CS-1.

\subsection{CS-3 type models}
Georgi and Machacek argued that, as long as the bosonic Lagrangian is
invariant under an $\rm SU(2)_L \times SU(2)_R$ symmetry, and all
scalar multiplets belong to $(N,N)$ representations for some $N$,
there will be a custodial symmetry.  In fact, this symmetry can be
realized in the CS-1 mode. We have seen that the scalar potential
might spoil this symmetry, forcing the model to be of the CS-2
variety, as it happens for 2HDM with the most general gauge invariant
potential.

As a matter of curiosity, we show here that CS-3 type models are also
possible.  In the SM, only one Higgs multiplet is needed to ensure CS.
There are higher representations where also just one multiplet
suffices.  One example is a 7-plet with $T=3$ and $n=-2$, whose
components are
\begin{eqnarray}
  \bm\Sigma = (\Sigma_5, \Sigma_4, \Sigma_3,\Sigma_2, \Sigma_1,
  \Sigma_0, \Sigma_{-1}) \,,
  \label{7plet}
\end{eqnarray}
where the subscripts indicate the electric charges.  Note that here
$\Sigma_{-1}$ is not the charge conjugate of $\Sigma_1$: they are
independent complex fields.  It is easily seen that a VEV of
$\Sigma_0$ would satisfy the condition of \Eqn{rho}, which means that
$\lag{mass}$ will have a CS.  So will $\lag{mixed}$, as has been
argued on general terms.  The unphysical Goldstone multiplet appearing
in $\lag{mixed}$ will involve 3 real fields out of the 14 that appear
in \Eqn{7plet}.  The real part of $\Sigma_0$ must be a CS-singlet.
But $\Sigma_5$ will have to be part of an 11-plet under CS, and there
are not enough fields left after setting aside the multiplets
mentioned already.  Therefore, we cannot have CS in other terms of
$\lag{kin}$. The same conclusion holds if the scalar sector holds only
a single multiplet higher than doublet, as there will not be enough neutral
fields to complete the CS multiplet.

Of course this multiplet is phenomenologically not interesting since
it cannot couple to fermions.  However, even after adding a doublet,
the CS in $\lag{mass}$ is maintained, but there is no $\rm SU(2)_L
\times SU(2)_R$ symmetry in the Lagrangian.

\section{The extended Georgi-Machacek Model} \label{s:eGM}
The GM model, obviously, implies degeneracy among the components of
the different multiplets of physical Higgs bosons.  In particular,
the masses of $F_{++}$, $F_+$ and $F_0$ turn out to be equal, and so
do the masses of $H_+$ and $H_0$, because of the CS.

However, one can have the CS in the CS-2 mode with the particle
content of the GM model.  This means $\rho=1$ but no degeneracy in the
scalar masses.  To our knowledge, this point has not been noticed in
the earlier literature.  This model will be called the extended
Georgi-Machacek model (eGM), and this is the most general model with
one SU(2) doublet, one real triplet, and one complex triplet scalar
leading to $\rho=1$. In the rest of the paper, we present this model
and discuss its phenomenological consequences.

\subsection{Requirements of the model}
To understand the existence of such a scenario, we start from the most
general gauge invariant potential given in \Eqn{gm.V16}.  With the
notation for the VEVs given in \Eqn{gm.vev} and the assumption that
they are all real, the minimization conditions of the potential are as
follows:
\begin{subequations}
\label{em.min}
\begin{eqnarray}
m_\phi^2 &=& -\mu_2 w -2 \surd2\mu_3 u + 2\lambda_\phi v^2
+ \left( \lambda_{\phi \chi} - \kappa_2\right)  u^2 + \lambda_{\phi \xi}
w^2 + 2\surd2 \kappa_3 uw\,, \\
u m_\chi^2 &=& \mu_1 u w - \surd2\mu_3 v^2 + 2\lambda_\chi u^3 +
\left(\lambda_{\phi \chi} - \kappa_2\right) v^2 u +  
\lambda_{\chi \xi}w^2 u + \surd2 \kappa_3 v^2 w\,,  \\
2 w m_\xi^2 &=& \mu_1 u^2-\mu_2 v^2+4\lambda_\xi w^3+2\lambda_{\phi
  \xi}v^2 w+2\lambda_{\chi \xi} u^2 w+2\surd2\kappa_3 v^2 u \,.
\end{eqnarray}
\end{subequations}
As said before, we need $u=w$ in order to obtain $\rho=1$.  Putting
$u=w$ in these equations and equating the coefficients of $u$, $u^3$,
$v^2$ and $uv^2$, we obtain the necessary conditions for obtaining the
equality of $u$ and $w$.  It turns out that these are just the conditions
given in \Eqn{gm.constr}.  The other conditions, shown
in \Eqn{gm.extra}, are not really necessary for obtaining $\rho=1$.

We can therefore consider a model with the scalar content of the GM
model, but where the most general Higgs potential is constrained only
by the conditions of \Eqn{gm.constr}.  This is what we call the
``extended GM model'' (eGM), and discuss in some detail.  This has 12
free parameters in the scalar potential, compared to 9 for the GM
model, as the three constraint conditions in \Eqn{gm.extra} are not
used.  This is a CS-2 type model.  Since the gauge sector is the same
as that of GM, it is obvious that $\lag{kin}$ is invariant under CS.
However, the scalar potential does not respect the symmetry.

  From a purely
gauge theoretical point of view, there is not much difference in the
symmetries of CS-1, CS-2 or CS-3 type models.  In all cases, the
symmetry is not obeyed by the entire Lagrangian with a realistic
fermion spectrum.  Thus, the symmetry is realized in a limited sector
of the theory, and is broken by loop corrections in all cases.  Since
the major consequence of CS, the mass relation of \Eqn{rho=1}, is
found to be consistent with experiments, one should explore different
scenarios in which this relation can be obtained.  This is what we do
in the rest of this paper.

\subsection{Mass terms in the general potential}
To understand what kind of masses will be observed for the physical
Higgs bosons, it is necessary to have a discussion of the masses that
arise from the most general gauge invariant potential, \Eqn{gm.V16},
assuming no equality among the VEVs.

First, we give the mass of the doubly charged scalar:
\begin{eqnarray}
  M_{++}^2 = -2 \mu_1 w +  \surd2 (\mu_3 - \kappa_3 w) v^2/u  + 4
  \tilde\lambda_\chi u^2 + 2 \kappa_2 v^2  \,. 
  \label{em.M++}
\end{eqnarray}
Next comes the mass squared matrix of the singly charged fields, in
the basis $\phi_+$-$\chi_+$-$\xi_+$:
\begin{eqnarray}
\mathbb M_+^2=
\begin{bmatrix}
\displaystyle{2\mu_2 w + 2\surd2 (\mu_3-\kappa_3 w)u \atop + 2 \kappa_2
  u^2} & -2\mu_3 v - \surd2 \kappa_2 uv & -\surd2\mu_2v + 2 \kappa_3 uv
\\[4mm] 
-2\mu_3 v - \surd2 \kappa_2 uv & \displaystyle{-\mu_1 w + \surd2(\mu_3-
  \kappa_3 w) \frac{v^2}{u} \atop + \kappa_1w^2 + \kappa_2v^2}
& \mu_1 u - \kappa_1 uw + \surd2 \kappa_3 v^2\\[4mm]
- \surd2 \mu_2 v + 2 \kappa_3 uv & \mu_1 u - \kappa_1 uw + \surd2\kappa_3
v^2 & \displaystyle{\kappa_1u^2 + \frac{1}{w}(-\mu_1 u^2 \atop
  +\mu_2 v^2-2\surd2\kappa_3 v^2 u)} \\  
\end{bmatrix}
\label{em.M+}
\end{eqnarray}

There are some instructive features of this mass matrix, which we
describe now.
\begin{enumerate}
\item There is a zero eigenvalue.  The corresponding eigenvector is
  proportional to $v\phi_+ + \surd2u\chi_+ + \surd2 w\xi_+$.  This is
  the unphysical charged Goldstone boson. Note that in the CS limit this
  combination reduces to the $h_+$ shown in \Eqn{gm.h}.
  
  \item If all the $\mu$-type and $\kappa$-type couplings are zero,
    $\mathbb M_+^2=0$, i.e, there are 3 singly-charged massless
    scalars. This is expected since the potential in this case has a
    $\rm [SU(2) \times U(1)]^3$ symmetry corresponding to separate
    transformations on the $\phi$, the $\chi$ and the $\xi$, and the
    VEVs break all of them. In fact, this is why our notation
    for using different letters for various quartic couplings is
    helpful: the masses in this sector do not depend on the
    $\lambda$-type couplings, as seen in \Eqn{em.M+}.

  \item Suppose only $\mu_1$ and $\kappa_1$ are nonzero among all
    $\mu$-type and $\kappa$-type couplings.  In this case, the
    potential has an $\rm [SU(2) \times U(1)]$ symmetry corresponding
    to the $\phi$ field only, and another  $\rm [SU(2) \times U(1)]$
    encompassing $\chi$ and $\xi$.  Sure enough, the mass matrix
    reduces to
    \begin{eqnarray}
      \mathbb M_+^2 = \begin{bmatrix}
        0 & 0 & 0 \\[6pt]
        0 & -\mu_1 w+\kappa_1 w^2 & \mu_1 u - \kappa_1 uw \\[6pt]
        0 & \mu_1 u - \kappa_1 uw & \kappa_1u^2- \mu_1 u^2/w
      \end{bmatrix} ,
      \label{em.case1}
    \end{eqnarray}
    which has two zero modes, $\phi_+$ and $u\chi_+ + w \xi_+$.

  \item Now suppose, among the $\mu$-type and $\kappa$-type couplings,
    only $\mu_3$ and $\kappa_2$ are non-zero. Now there is an
    independent SU(2) for $\xi$, and
    \begin{eqnarray}
      \mathbb M_+^2=
      \begin{bmatrix}
        2\surd2 \mu_3 u + 2\kappa_2 u^2 & -2 \mu_3 v - \surd2\kappa_2 uv
        & 0 \\[6pt] 
        -2 \mu_3 v - \surd2\kappa_2 uv & \surd2 \mu_3 v^2/u + \kappa_2
        v^2 & 0\\[6pt] 
        0 & 0 & 0
      \end{bmatrix}\,.
      \label{em.case2}
    \end{eqnarray}

  \item Similarly, if only $\mu_2$ is non-zero, $\chi_+$ will be
    massless, and so will be the combination $v\phi_++\surd2w \xi_+$. 
    
\end{enumerate}

There are two CP-odd neutral scalar fields: $\phi_0''$ and
$\chi_0''$.  Their mass matrix, choosing the basis in that order, is:
\begin{eqnarray}
  \mathbb M_{\rm odd}^2 = \surd2 (\mu_3 - \kappa_3 w)
  \begin{bmatrix}
    4 u & -2v \\
    -2v & v^2/u
  \end{bmatrix}
  \label{em.Modd}
\end{eqnarray}
Since real fields are involved here, the actual terms in the
Lagrangian come with a factor of $1/2$.  The zero eigenmode is $v
\phi_0'' + 2 u \chi_0''$, which is the unphysical Goldstone mode shown
in \Eqn{gm.h0}.  In various limits discussed in context of the singly
charged scalar matrix, the behaviour of this matrix also follows the
pattern expected from symmetry.

Finally, there are the CP-even neutral fields, the real parts of the
neutral component of each multiplet.  With the general 16-parameter
potential and different VEVs as indicated in \Eqn{gm.vev}, these terms
are as follows, in the basis $\phi_0'$-$\chi_0'$-$\xi_0'$: 
\begin{eqnarray}
\mathbb M_{\rm even}^2=
\begin{bmatrix}
4 \lambda_\phi v^2 
& \displaystyle{2\surd2 ({-}\mu_3 + \kappa_3 w) v \atop
+ 2( -\kappa_2 + \lambda_{\phi\chi}) uv}
& \displaystyle{- \surd2 \mu_2v + 4 \kappa_3 uv \atop + 
   2\surd2 \lambda_{\phi\xi} vw}
\\[8mm]
\displaystyle{2\surd2 ({-}\mu_3 + \kappa_3 w) v \atop
+ 2( -\kappa_2 + \lambda_{\phi\chi}) uv}
& \displaystyle{ (\mu_3
  - \kappa_3w) \frac{\surd2v^2}{ u} \atop + 4 \lambda_\chi u^2} 
& \displaystyle{ \surd2 \mu_1 u \atop
  + 2\surd2 \lambda_{\chi\xi} uw + 2\kappa_3 v^2} 
\\[8mm]
\displaystyle{- \surd2 \mu_2v  + 4 \kappa_3 uv \atop + 
   2\surd2 \lambda_{\phi\xi} vw}
& \displaystyle{ \surd2 \mu_1 u \atop
  + 2\surd2 \lambda_{\chi\xi} uw + 2\kappa_3 v^2} 
& \displaystyle{ ({-} \mu_1u^2 + \mu_2 v^2  - 2\surd2 \kappa_3
v^2u){1 \over w} \atop +
8 \lambda_\xi w^2}
\end{bmatrix}.
\label{em.Meven2}
\end{eqnarray}
We will discuss the masses that come out of these expressions after
putting the constraints applicable to the eGM model.

\subsection{Masses in the eGM model}
In the eGM model, we impose the four constraints
of \Eqn{gm.constr}, which ensure $u=w$, as shown earlier.  We will
henceforth denote the physical states by a prime, so for the doubly
charged scalar, $F_{++} = F'_{++}$. Its mass is obtained
from \Eqn{em.M++} by putting the equality of the two VEVs:
\begin{eqnarray}
  M_{++}^2 = -2 \mu_1 u + \surd2 \mu_3  v^2/u  
  + 4 \tilde\lambda_\chi u^2 + (2 \kappa_2 - \surd2 \kappa_3) v^2 \,. 
  \label{em.mass++}
\end{eqnarray}
The mass matrix of the singly charged scalars was given
in \Eqn{em.M+}, where we now put $u=w$. In the basis spanned by 
$h_+$-$H_+$-$F_+$ defined in \3Eqs{gm.h} {gm.H} {gm.F},
the mass matrix reduces to
\begin{eqnarray}
\mathbb M_+^2 = \frac{\surd2}{\sin 2\beta} \begin{bmatrix}
0 & 0 & 0 \\
0 & (A-C) &  (A+C)\, \cos\beta \\ 
0 & (A+C)\, \cos\beta & (A-C)\, \cos^2\beta + \surd2B\, \sin 2\beta \\ 
\end{bmatrix} \,, 
\label{em.blockM+}
\end{eqnarray}
where $\tan\beta = 2u/v$ as defined earlier, and we have used the
shorthands
\begin{subequations}
\label{em.ABC}
\begin{eqnarray}
A &=& 2\mu_3 v + \surd2 \kappa_2 uv \,, \\
B &=&  -\mu_1 u + \kappa_1 u^2 - \surd2 \kappa_3 v^2 \\
C &=& -2\mu_3v + 2 \kappa_3 uv \,.
\end{eqnarray}
\end{subequations}
The zeroes in the first row and the first column are not surprising:
they just confirm that $h_+$ is indeed a massless mode.  The other two
eigenvectors are
\begin{eqnarray}
H'_+ &=& \cos\theta_+ H_+ - \sin\theta_+ F_+ \,, \nonumber\\
F'_+ &=& \sin\theta_+ H_+ + \cos\theta_+ F_+ \,, 
\label{em.H'F'}
\end{eqnarray}
with
\begin{eqnarray}
\tan 2\theta_+ = {2(A+C)\, \cos\beta \over
\surd2 B\, \sin 2\beta - (A-C)\sin^2\beta} \,.
\end{eqnarray}

We now discuss the neutral fields. There is only one CP-odd neutral
scalar, which we call $H'_0$. It is easily seen, from \Eqn{em.Modd},
that this is the combination proportional to $2u \phi_0'' -
v \chi_0''$, and its mass is given by
\begin{eqnarray}
M_{\rm odd}^2 = \surd2 (\mu_3 - \kappa_3u) (4u^2 + v^2)/u \,,
\label{em.oddmass}
\end{eqnarray}
the only non-zero eigenvalue of the matrix.

It is instructive to see the GM limit from these results.  Note
that \Eqn{em.ABC} tells us that $A+C=(\surd2\kappa_2+2\kappa_3)uv$,
which vanishes in the GM limit, as seen from \Eqn{gm.c5}.  Thus
$\theta_+=0$, implying that $H_+$ and $F_+$ are truly the mass
eigenstates.  Moreover, note that in this limit, the mass square of
$H_+$ is the middle diagonal element of the matrix
of \Eqn{em.blockM+}, which is
\begin{eqnarray}
\frac{ \surd2} {\sin 2\beta} \,  (A-C) = \surd2 (\mu_3 - \kappa_3u) (4u^2 +
v^2)/u \,.  
\end{eqnarray}
This is equal to the value that appears in \Eqn{em.oddmass}, which is
expected since $H_\pm$ and $H_0$ form a CS triplet.  However, in the
eGM, this equality is not maintained, because of extra terms that
appear for $\theta_+ \neq 0$.

Similarly, we note from \Eqn{em.blockM+} that $(\mathbb M_+^2)_{33}$
is given by
\begin{eqnarray}
\frac{\cot\beta}{\surd2}\, (A-C) + 2B = -2\mu_1 u + \surd2 \mu_3 v^2/u
+ 2\kappa_1 u^2 
+ \frac12 ( \kappa_2 - 5\surd2 \kappa_3 ) v^2 \,.
\label{em.massF+}
\end{eqnarray}
If we put the extra GM conditions given in \Eqn{gm.extra}, this
expression equals the expression given in \Eqn{em.mass++} for the
doubly charged Higgs boson, which happens because in GM model the
fields are part of a degenerate 5-plet.

Finally, let us discuss the masses of the CP-even neutral scalars.  In
the basis $\phi_0'$-$\chi_0'$-$\xi_0'$ (which are defined
in \Eqn{vev-define}, and not to be confused with the primed mass
eigenstates), the mass terms appearing in the Lagrangian can be
written in the form of the matrix
\begin{eqnarray}
\mathbb M_{\rm even}^2=
\begin{bmatrix}
4 \lambda_\phi v^2 
& \displaystyle{-2\surd2 \mu_3 v \atop + 4( \lambda_{\phi\xi} + \surd2
  \kappa_3) uv} 
& \displaystyle{- 2 \mu_3 v \atop + 
   (2\surd2 \lambda_{\phi\xi} + 4 \kappa_3) uv}
\\[8mm]
\displaystyle{-2\surd2 \mu_3 v \atop + 4( \lambda_{\phi\xi} + \surd2
  \kappa_3) uv} 
& \displaystyle{ 4 \lambda_\chi u^2 + (\mu_3
  - \kappa_3u) \frac{\surd2v^2}{ u} } 
& \displaystyle{ \surd2 \mu_1 u + 2\kappa_3 v^2 \atop
  + 4\surd2 (\lambda_\chi - 2\lambda_\xi) u^2 } 
\\[8mm]
\displaystyle{- 2 \mu_3 v \atop + 
   (2\surd2 \lambda_{\phi\xi} + 4 \kappa_3) uv}
& \displaystyle{ \surd2 \mu_1 u + 2\kappa_3 v^2 \atop
  + 4\surd2 (\lambda_\chi - 2\lambda_\xi) u^2 } 
& \displaystyle{ - \mu_1u + \mu_2 {v^2 \over u} \atop +
8 \lambda_\xi u^2 - 2\surd2 \kappa_3 v^2 }
\end{bmatrix}
\label{em.Meven}
\end{eqnarray}
We have put the restriction $u=w$, as well as the constraints shown in
\Eqn{gm.constr} that are necessary for obtaining the equality of the
VEVs, thereby eliminating the couplings $\mu_2$, $\kappa_2$, and
$\lambda_{\chi\xi}$.  But we have not used the conditions of
\Eqn{gm.extra}.

It can be easily seen that the combination $F_0$, shown in \Eqn{gm.F},
is an eigenvector of this matrix, and the eigenvalue is
\begin{eqnarray}
M_{F_0}^2 = {-}2 \mu_1u + \surd2 \mu_3 {v^2 \over u} + 4 (4\lambda_\xi
- \lambda_\chi) u^2 - 3 \surd2 \kappa_3 v^2 \,,
\label{em.massF0}
\end{eqnarray}
so that the mass of $F_0$, in the GM limit with \Eqn{gm.extra}, is the
same to that of $F'_+$ and $F_{++}=F'_{++}$, confirming the degeneracy
of the 5-plet of CS for the GM model.

It is important to note that there was no need for using the
constraints of \Eqn{gm.extra} in order to show that $F_0$ is an
eigenstate of the mass matrix.  It means that this conclusion is valid
for eGM as well, in presence of some extra terms in the potential
which were not present in the GM model.  The physics of the eGM model
can therefore be better understood if we write the matrix
of \Eqn{em.Meven} in the basis $S_1$-$S_2$-$F_0$, defined
in \Eqs{gm.F}{gm.S}.  We impose the conditions of \Eqn{gm.constr} on
the couplings which are necessary for obtaining $\rho=1$, and the 
resulting matrix is
\begin{eqnarray}
\mathbb M_{\rm even}'^2 = \begin{bmatrix}
4 \lambda_\phi v^2 
& \displaystyle{-2\surd3 \mu_3v + 2\surd6 \lambda_{\phi\xi} uv \atop +
4\surd3 \kappa_3 uv} 
& 0
\\[8mm]
\displaystyle{-2\surd3 \mu_3v + 2\surd6 \lambda_{\phi\xi} uv \atop +
4\surd3 \kappa_3 uv} 
& \displaystyle{\mu_1 u + \surd2\mu_3 {v^2 \over u} \atop -
8\lambda_\xi u^2 + 8 \lambda_\chi u^2 }
& 0
\\[8mm]
0
& 0
& \displaystyle{-2\mu_1u + \surd2 \mu_3 {v^2 \over u} \atop
+ 4(4\lambda_\xi - \lambda_\chi) u^2 - 3\surd2 \kappa_3 v^2} 
\end{bmatrix}\,.
\label{em.Mevenblock}
\end{eqnarray}

So the eigenstates of this matrix may be called $S'_1$, $S'_2$ and
$F'_0 = F_0$, where the first two are combinations of the CS singlets
shown in \Eqn{gm.S}:
\begin{eqnarray}
\begin{pmatrix} S'_1 \\ S'_2 
\end{pmatrix} =
\begin{pmatrix} \cos \theta_0 & -\sin \theta_0 \\
\sin\theta_0 & \cos\theta_0
\end{pmatrix}
\begin{pmatrix} S_1 \\ S_2 
\end{pmatrix} \,, 
\end{eqnarray}
where
\begin{eqnarray}
\tan 2\theta_0 = {2(-2\surd3 \mu_3v + 2\surd6 \lambda_{\phi\xi} uv +
4\surd3 \kappa_3 uv) \over \mu_1 u + \surd2\mu_3 {v^2 \over u} -
8\lambda_\xi u^2 + 8 \lambda_\chi u^2 - 4 \lambda_\phi v^2} \,.
\end{eqnarray}
For small values of the mixing angle $\theta_0$, $S'_1$ is dominantly
the doublet and we expect its mass to be around 125 GeV.


\section{Constraints on the eGM potential}{\label{sec:RGE}}
Once we establish the eGM model as a more general extension of the
conventional GM model with $\rho=1$ at the tree-level ensured, one may
now look at the possible phenomenological signatures.  Similar
signatures for a CS-conserving general Higgs sector have been
discussed in Ref.~\cite{Kundu:1995qb}.  For our case, however, most of
the signatures are quite similar to the GM model, like the scalar
production through gauge boson fusion mechanism, or ${\sf S}_1 \to
{\sf S}_2 {\sf V}$ and ${\sf S}\to {\sf V}_1{\sf V}_2$ decay channels
(where ${\sf S}$ stands for any generic scalar and ${\sf V}$ for any
generic gauge boson).  There exist a few notable differences, which we
mention here, and plan to explore later in more detail.

To set some benchmark points, we first need to enlist the constraints
on the potential. The theoretical constraints are quite standard, and
may be summarized as follows:
\begin{itemize}
\item The potential must be bounded from below, {\em i.e.}, there is
  no direction in the potential space that leads to an unbounded
  minimum. The constraints, for the most general 16-parameter model
  given by Eq.\ (\ref{gm.V16}), are as follows:
\allowdisplaybreaks  
\begin{subequations}
\label{stability}
\begin{eqnarray}
&& \lambda_\phi\,, \lambda_\xi\,, \lambda_\chi\,,
  \lambda_\chi+\tilde{\lambda}_\chi  > 0\,,\\ 
&& 2\sqrt{\lambda_\phi \lambda_\xi}+\lambda_{\phi \xi} > 0\,,\\
&& 2\sqrt{\lambda_\phi \lambda_\chi}+\lambda_{\phi \chi} > \left\vert
  \kappa_2  \right\vert\,, \\ 
&& 2\sqrt{\lambda_\phi(\lambda_\chi+\tilde{\lambda}_\chi)}+\lambda_{\phi \chi} > 0\,,\\
&& \lambda_{\chi\xi} > -2\sqrt{ \lambda_\chi \lambda_\xi } \qquad  \mbox{for $\tilde{\lambda}_\chi > 0$, $\kappa_1 > 0$},\\
&& \lambda_{\chi\xi} > -2\sqrt{ \lambda_\chi \lambda_\xi } +
  \frac{1}{2}\left\vert\kappa_1\right\vert \qquad \mbox{for
    $\tilde{\lambda}_\chi > 0$, $\kappa_1 < 0$}, \\
&& \lambda_{\chi\xi} > -2\sqrt{ (\lambda_\chi + \tilde{\lambda}_\chi)
    \lambda_\xi } \qquad \mbox{for $\tilde\lambda_\chi < 0$, $\kappa_1
    > 0$},\\ 
&& \lambda_{\chi\xi} > -2\sqrt{ (\lambda_\chi + \tilde{\lambda}_\chi)
    \lambda_\xi } + \left\vert\kappa_1\right\vert \qquad \mbox{for
    $\tilde\lambda_\chi < 0$, $\kappa_1 < 0$}. 
\end{eqnarray} 
\end{subequations}
This shows that apart from $\lambda_\phi$, $\lambda_\chi$ and
$\lambda_\xi$, none of the quartic couplings need to be positive. The
constraint on $\kappa_3$ is more complicated, but once we go to the CS
limit by imposing the relations shown in \Eqn{gm.constr}, $\kappa_3$
gets related with $\lambda_{\phi\chi}$, $\lambda_{\phi\xi}$, and
$\kappa_2$. Also, in the CS limit, one finds $\lambda_\chi >
\lambda_\xi$ if $\tilde{\lambda}_\chi, \kappa_1 > 0$, and for all
other cases one has to replace $\lambda_{\chi\xi}$ with $2\lambda_\chi
- 4\lambda_\xi$.

\item The partial wave amplitudes for any scattering process must not
  violate the unitarity bounds: 
\begin{eqnarray}
\left\vert {\rm Re}(a_\ell)\right\vert \, \leq\, \frac12\,,
\end{eqnarray}
where $a_\ell$ is the $\ell$-th partial wave amplitude in any
channel. This leads to the partial wave unitarity constraints, shown
for the conventional GM model in Refs.\ \cite{Aoki:2007ah,
  Hartling:2014zca}.  While we deviate from the conventional GM model
by not imposing the conditions of \Eqn{gm.extra}, quartic couplings
with magnitudes of the order of unity are expected to saturate the
unitarity bounds.

\item The oblique parameters, particularly $S$ and $T$, are within the
  experimental limits. This is expected, as $\alpha T = \rho -1$ is
  already zero at the tree-level, and $S$ receives only logarithmic
  corrections from scalars.

\item One must also ensure that the Landau pole is not hit too soon,
  or, in other words, none of the couplings blow up at a sufficiently
  low energy. We would like to demand all the couplings to be
  well-behaved below 
  40 
  TeV. To be rigorous, since we use one-loop
  renormalization group equations (RGE) to calculate the evolution,
  the demand is on the perturbative nature of the couplings. The RGEs
  are taken from Ref.\ \cite{Keeshan:2018ypw} and displayed in
  Appendix \ref{app.rge}.  For our benchmark points, and indeed over
  almost the entire parameter space, the potential never becomes
  unstable with RGE evolution.
  
\item We have also ensured that over the allowed parameter space, all
  the scalar mass eigenstates are positive.

\end{itemize}

The experimental bounds on the scalar masses depend on the triplet
VEV, and may be found in Refs.\ \cite{CMS:2017fgp, CMS:2021wlt,
  Ismail:2020kqz}.  However, one may note that these are bounds in the
GM model. We discuss below where eGM differs from GM.

\begin{table}[t]
\begin{center}
\begin{tabular}{||c||c|c|c|c|c|c||}
\hline
Benchmark &  $\mu_1$  &  $\mu_3$  &  $\kappa_1$ & $\kappa_2$ &
$\kappa_3$ & $\lambda_\phi$ \\ 
\hline
BP-1 & $-9.13$ & 29.03 & 0.12 & $-1.41$ & $-1.09$ & 0.140  \\
BP-2 & $-2.56$ & 0.55 & 0.50 & 3.80 & $-3.10$ & 0.134 \\
\hline
Benchmark & $\lambda_\chi$ & $\vphantom{q^2 \over q}\tilde{\lambda}_\chi$ &  $\lambda_{\phi\chi}$ & $\lambda_{\chi\xi}$ & $v$ & $u$ \\
\hline
BP-1 & 0.76 & 1.19 & 1.78 & 0.45 & 173 & 10 \\
BP-2 & 1.50 & 0.70 & 3.85 & 0.10 & 174 & 1 \\
\hline
\end{tabular}
\caption{The benchmark points, with $\mu_1$, $\mu_3$, $u$ and $v$ in GeV.}
\label{tab:bench}
\end{center}
\end{table}

\section{A cursory analysis}{\label{sec:results}}
In this Section, we would like to highlight some crucial differences
between the GM and the eGM models.  For this, we choose two benchmark
points, which have been dubbed BP-1 and
BP-2 in Table~\ref{tab:bench}. The points have been chosen in a way that
they satisfy the constraints enlisted in Section \ref{sec:RGE}. 
While BP-1 has been chosen to highlight the differences between GM and eGM, 
BP-2 has been chosen in such a way as to reproduce some hints of new 
scalars found at the LHC, like a neutral scalar at 151 
GeV, a pseudoscalar at about 400 GeV, and another heavy scalar at 
about 660 GeV~\cite{Richard:2021edf,Crivellin:2021ubm,Richard:2021ovc}.
In addition, we put some extra conditions, based on the following
considerations.  
\begin{enumerate}
\item The lightest CP-even scalar is at 125 GeV, and is dominantly a
  doublet. To ensure consistency with the LHC data, we demand that the
  doublet component is at least 90\%. Note that triplet admixture
  suppresses the branching fractions to the fermionic channels, and
  enhances the same to the di-gauge channels.

\item The model must be valid till, at least, 
40
 TeV. This is a rather
conservative estimate, even a 10 TeV upper limit would have easily passed all
the experimental constraints. In other words,
  none of the couplings should blow up below that energy scale. This
  ensures that the pure quartic couplings $\lambda_\phi$,
  $\lambda_\chi$, and $\lambda_\xi$ cannot have large positive values
  to start with, at the electroweak scale. 
  In fact, for BP-1, the model remains well-behaved well beyond 100 TeV, while for BP-2, 
  the Landau pole is hit just beyond 40 TeV, because of the large values of the couplings that 
  we have started with.
  We also find, as expected,
  that the $\rho$ parameter deviates from unity at a high scale (as
  the VEVs are functions of couplings which are in turn
  energy-dependent), but that is not a serious concern as it has been
  and will be measured only at the electroweak scale. The GM model
  keeps $\rho=1$ at all energy scales, with the well-known caveat of
  divergent oblique parameters when the U(1)$_{\rm Y}$ part is gauged.

\item If the doubly charged scalar $F'_{++}$ is below 1 TeV, the
  triplet VEV cannot be too large, or that would result in an
  unacceptably large rate for $F'_{++}\to W_+W_+$ at the LHC. We have
  used the constraint on the triplet VEV from
  Ref.\ \cite{Ismail:2020kqz}. The strongest bound comes from the CMS
  search on the channel mentioned above, which roughly translates to
  $u=w < 18$ GeV for the average 5-plet mass around 300 GeV. This
  bound still holds in eGM as $F'_{++}$ has no doublet admixture.

\end{enumerate}

 \begin{table}[b]
\begin{center}
\begin{tabular}{||c|c||c|c||c|c||}
\hline
\multirow2*{Type} & \multirow2*{Field} 
& \multicolumn{2}{c||} {BP-1} & \multicolumn{2}{c||} {BP-2} \\
\cline{3-6}
 &  & Mass (GeV) & $\phi$-comp. & Mass (GeV) & $\phi$-comp. \\
\hline
\multirow3*{CP-even neutral}   & $S'_1$ & 125 & 0.995 & 125 & 0.964 \\
 & $S'_2$  & 352 & $0.097$& 155 & $0.265$ \\
  & $F'_0$ & 512 & 0 & 649 & 0 \\
  \hline
  CP-odd neutral & $H'_0$ & 414 & 0.115 & 395 & 0.011 \\
  \hline
\multirow2*{Singly charged} & $F'_+$ & 329 & $0.106$ & 643 & $4\times 10^{-4}$ \\
                           & $H'_+$ & 486 & $0.043$ & 384 & $0.011$  \\
  \hline
  Doubly charged & $F'_{++}$ & 293 & 0 & 622 & 0 \\
  \hline 
\end{tabular}
\caption{Masses of various scalars, and their doublet admixture, for
  the two benchmark points chosen in Table~\ref{tab:bench}. By
  $\phi$-component of any physical field $X'$, we mean $\left\vert\bra
  X' | \phi\ket\right\vert$.}
\label{tab:mass}
\end{center}
\end{table}

The choice of the benchmark points ensure that $M_W =80.41$ GeV, and the
lightest CP-even scalar $S'_1$ remains at 125 GeV, with at least 95\%
doublet component.
All the other theoretical and experimental constraints have also been
checked. Note that the two input parameters $m_\phi^2$ and $m_\chi^2$
have been replaced by the VEVs $v$ and $u$ respectively.

For these two benchmarks, the scalar masses and the doublet ($\phi$)
component in each of them are shown in Table \ref{tab:mass}. Only the
physical states are displayed, which includes three CP-even neutral,
one CP-odd neutral, two pairs of singly charged, and one pair of
doubly charged scalars.

\begin{figure}
  \begin{center}
    \includegraphics[page=1, width=80mm, clip,
      trim=50mm 160mm 65mm 40mm] {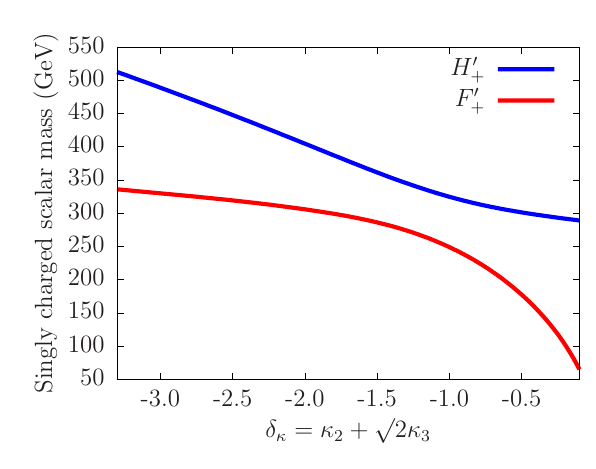} \hfill
    \includegraphics[page=2, width=80mm, clip,
      trim=50mm 160mm 65mm 40mm] {allfigs.pdf}
  \end{center}
  
\caption{\small Left: The variation of the physical singly charged
  scalar masses, $M_{H'_+}$ (blue) and $M_{F'_+}$ (red), with
  $\delta_\kappa = \kappa_2 + \surd2\kappa_3$, which is zero at the GM
  limit. Only $\kappa_3$ has been varied, and $\lambda_\phi$ suitably
  adjusted to keep $m_{S'_1}=125$ GeV, with all other parameters of
  the scalar potential fixed at their BP-1 values.  Right: The
  variation of the doublet (leading to fermionic decay modes)
  component admixture to $H'_+$ (blue) and $F'_+$ (red). Again,
  $\kappa_3$ has been varied. Note that in the GM model, $F'_+$ does
  not have any doublet component.}
\label{fig:split}
\end{figure} 
In Fig.\ \ref{fig:split}, we show how the masses of the physical
singly charged scalars, $H'_+$ and $F'_+$, vary with
$\delta_\kappa\equiv \kappa_2+\surd2 \kappa_3$, which is zero in the
GM limit. All the parameters of the scalar potential are fixed at the
benchmark point BP-1, except $\kappa_3$, and $\lambda_\phi$ has been
slightly adjusted to keep the mass of one of the CP-even neutral
scalars, $S'_1$, at 125 GeV.  The lower limit of $\delta_\kappa$ is
kept at about $-\sqrt{4\pi}$, whereas the upper limit ($\sim -0.1$) is
where the potential develops a saddle point instead of a true local
minimum; the mass squared terms of the 5-plet scalars become
negative. We also show the doublet admixture in $F'_+$ and $H'_+$ as a
function of $\delta_\kappa$. Note that in the GM limit, $F'_+$ does
not have any doublet component. The crossover, as can be seen in the
right-hand panel of Fig.\ \ref{fig:split}, occurs at $\kappa_3=0$.

\begin{figure}
    \begin{center}
    \includegraphics[page=3, width=55mm, clip,
      trim=55mm 162mm 67mm 45mm] {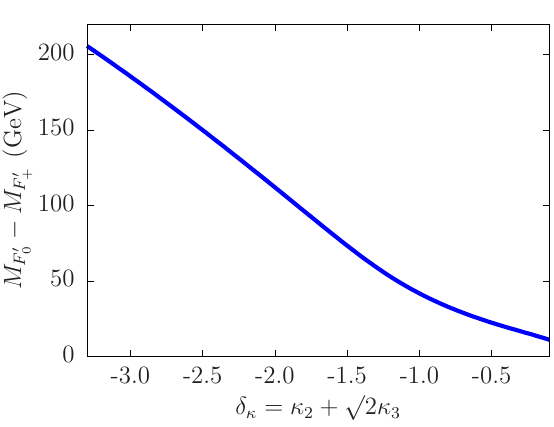} \hfill
    \includegraphics[page=4, width=55mm, clip,
      trim=55mm 162mm 67mm 45mm] {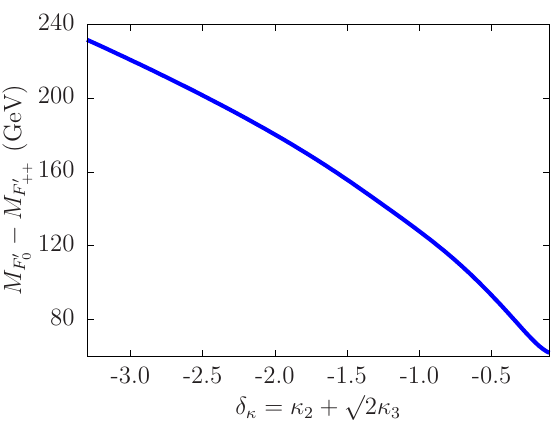} \hfill
    \includegraphics[page=5, width=55mm, clip,
      trim=55mm 162mm 67mm 45mm] {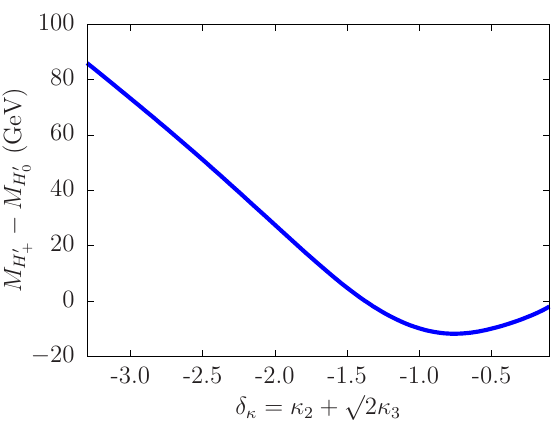}
  \end{center}

\caption{\small{Left and center: The splitting among the members of
    the 5-plet as a function of $\delta_\kappa$. The definition of
    $\delta_\kappa$, as well as its variation, is the same as in
    Fig.\ \ref{fig:split}. Right: The same for the members of the
    physical triplet.}}
\label{fig:doub}
\end{figure}

In Fig.\ \ref{fig:doub}, we show the mass splitting among the members
of the triplet and those of the 5-plet, as a function of
$\delta_\kappa$, again varied around BP-1 as mentioned before.  The
nature of the plots changes if we vary one or more of the other
parameters, or probe the violation of other GM constraints shown in
\Eqn{gm.extra}, but such a detailed scan is not our goal.

Because of these features, the eGM model has a richer phenomenology,
as commented earlier.  Here, we list some of the new consequences to
be expected from the eGM model, without going into a detailed
analysis.

\begin{itemize}

\item The members of the physical scalar multiplets are not
  degenerate. This leads to the decays like $F'_0 \to F'_+ W^*$, $F'_+
  \to F'_{++}W^*$, $H'_0 \to H'_+ W^*$, and their CP-conjugate modes.
  If the splitting is not too large, the virtual $W$ will either lead
  to a soft pion or to a pair of soft jets or leptons.  However, the
  final states should be observable at the LHC if the boost is
  significant.  On the other hand, for typical benchmark points as
  shown above, the splittings are large, and one may expect isolated
  hard leptons, over most of the parameter space.  How to observe this
  signal over the SM background is a different issue.

\item The neutral and singly-charged scalars belonging to the physical
  triplet and 5-plet contain some nontrivial component of the SU(2)
  doublet $\phi$. This leads to fermion-antifermion final states that
  never come from pure SU(2) triplets. The production cross-section
  will, of course, be suppressed by the mixing angle squared, and such
  signals may be interesting only at HL-LHC.
  
\item
Hints of several scalar resonances have been found at the LHC, {\em e.g.}, 
a neutral scalar at 151 GeV, a pseudoscalar close to 400 GeV, and a heavy scalar
at about 660 GeV \cite{Richard:2021edf}. BP-2 shows that it is possible to 
generate the spectrum in eGM. However, a detailed collider study, matching 
theoretical predictions with corresponding signal strengths, lies beyond the
scope of this paper.

\end{itemize}

\section{Conclusion} \label{sec:conclude}
In this paper, we critically discuss the concept of custodial
symmetry. Defining it as something that keeps the SU(2) gauge bosons
mass degenerate (in the limit $\cos\theta_W=1$), we classify CS in
three categories, depending on which parts of the action remain
invariant under this. For example, we explicitly demonstrate that
while the SM action respects CS, it is not the case for the most
general 2HDM.

After this groundwork, we focus our attention to the Georgi-Machacek
model with one scalar doublet and two triplets, which, by
construction, keeps $\rho=1$. The phenomenology and collider
signatures of the GM model have been well explored, so we try to find
whether this is the most general model with such scalar multiplet
assignment to respect CS.  We realize that this is not so; one may
extend this model by introducing three more parameters in the
potential and still keep $\rho=1$ at the tree-level. This, of course,
changes the type of CS that the GM model respects. Instead of 9 for
the GM model, the scalar potential of this extended model, which we
call extended GM or eGM, is specified by 12 independent parameters and
we show that this is the most general extension of the scalar sector
with one doublet and two triplets keeping CS intact.

This, in turn, affects the corresponding phenomenology. While our
focus in this paper is not on the collider signals, we point out two
important differences between GM and eGM models that should be
relevant for a collider study. First, the CS multiplets, the triplet
and the 5-plet, are no longer degenerate in the eGM model, and
secondly, both the physical singly charged scalars have a doublet
component which leads to a fermion-antifermion final state, and hence
may change the search strategies.  We will take up this study in
detail in a subsequent publication.

\bigskip



{\bf Acknowledgement}:
We acknowledge helpful conversation with
Biplob Bhattacherjee. 
We also thank Fran\c{c}ois Richard for helpful comments 
on the first version of the manuscript and sharing some interesting
results with us.
AK has been supported by the Science and Engineering Research
Board, Government of India, through the
grants CRG/2019/000362, MTR/2019/000066, and
DIA/2018/000003. PBP thanks the same agency
for the grant EMR/2017/001434.

\appendix
\renewcommand{\theequation}{\thesection.\arabic{equation}}
\makeatletter
\@addtoreset{equation}{section}
\makeatother


\appendix

\section{SU(2) invariants in spherical basis}\label{app:sb}
When we write the components of an SU(2) multiplet,  {\em e.g.},
in \Eqn{gm.fields}, we usually employ the ``spherical'' notation,
meaning that each component has a well-defined value of $T_3$, the
diagonal generator of SU(2).  When we write the SU(2) generators, on
the other hand, we tend to use the Cartesian ones.  This mixed
notation was employed in writing the GM potential in the
literature \cite{Georgi:1985nv, Logan:2015xpa}.  In this paper, we
have used a spherical notation for everything.  In this Appendix, we
show the correspondence between the two notations and explain the way
we write \Eqn{gm.pot}, for example.

In terms of the Cartesian components, the SU(2) generators in the
3-dimensional representation are as follows:
\begin{eqnarray}
t_x = \begin{pmatrix}
0 & 0 & 0 \\
0 & 0 & -i \\
0 & i & 0
\end{pmatrix} \,, \qquad
t_y = \begin{pmatrix}
0 & 0 & i \\
0 & 0 & 0 \\
-i & 0 & 0
\end{pmatrix} \,, \qquad
t_z = \begin{pmatrix}
0 & -i & 0 \\
i & 0 & 0 \\
0 & 0 & 0 
\end{pmatrix} \,.
\label{sb.txyz}
\end{eqnarray}
We can easily find a unitary matrix $U$ such that
$Ut_zU^\dagger \equiv t_0$ is diagonal.  This matrix is
\begin{eqnarray}
U = {1 \over \surd2}
\begin{pmatrix}
-1 & i & 0 \\
0 & 0 & \surd2 \\
1 & i & 0 
\end{pmatrix} \,,
\label{sb.U}
\end{eqnarray}
and this gives
\begin{eqnarray}
t_0 = \, \mathrel{\rm diag} (1, 0, -1) \,.
\label{sb.t0}
\end{eqnarray}
So, for a vector $A$, the spherical components and the Cartesian
components would be related by
\begin{eqnarray}
A_{\rm sph} = UA_{\rm Car} \,,
\end{eqnarray}
which gives
\begin{eqnarray}
A_{(+)} &=& {1 \over \surd2} (-A_x + iA_y) \,, \nonumber\\
A_{(0)} &=& A_z \,, \nonumber\\
A_{(-)} &=& {1 \over \surd2} (A_x + iA_y) \,.
\label{sb.+0-}
\end{eqnarray}
Note that the indices in parentheses denote the $T_3$-eigenvalues of
the components.  This can differ from the electric charge when the
U(1) quantum number is non-zero, {\em e.g.}, for the complex triplet
$\chi$ .  When we write the charges, we do not use the parentheses.
For the generators also, we do not use the parentheses since there is
no scope of confusion there.  The invariant combination of the two
vectors will be given by
\begin{eqnarray}
\bm A \cdot \bm B &=& A_x B_x + A_y B_y + A_z B_z \nonumber\\
&=& {-}A_{(+)} B_{(-)} - A_{(-)} B_{(+)} + A_{(0)} B_{(0)} \,.
\label{sb.dotprod}
\end{eqnarray}
The minus signs might look odd in this notation, but they are exactly
in correspondence with the Clebsch-Gordan coefficients obtained while
constructing a spin-0 combination from two spin-1 particles.

In the spherical notation, then, we should take the generators as
$t_0$ given in \Eqn{sb.t0}, and
\begin{eqnarray}
t_+ = {1\over \surd2} U(-t_x+it_y)U^\dagger &=&  
\begin{pmatrix}
0 & 0 & 0 \\
-1 & 0 & 0 \\
0 & -1 & 0 
\end{pmatrix} \,,  \nonumber\\
t_- = {1\over \surd2} U(t_x+it_y)U^\dagger &=&  
\begin{pmatrix}
0 & 1 & 0 \\
0 & 0 & 1 \\ 
0 & 0 & 0 
\end{pmatrix} \,, 
\label{sb.t+-}
\end{eqnarray}
where $t_x$ and $t_y$ have been read from \Eqn{sb.txyz}.  For the
2-dimensional representation, similarly, we will use the matrices
\begin{eqnarray}
\tau_+ &=&  {1\over \surd2} (-\tau_x + i \tau_y) =
\begin{pmatrix}
0 & 0 \\ -\surd2 & 0 
\end{pmatrix} \,, \nonumber\\
\tau_0 &=&  \tau_z = 
\begin{pmatrix}
1 & 0 \\ 0 & -1 
\end{pmatrix} \,, \nonumber\\
\tau_- &=&  {1\over \surd2} (\tau_x + i \tau_y) =
\begin{pmatrix}
  0 & \surd2 \\ 0 & 0 
\end{pmatrix} \,, 
\label{sb.tau}
\end{eqnarray}
where $\tau_x$, $\tau_y$ and $\tau_z$ are the usual Pauli matrices.
Although the SU(2) generators are really half of the Pauli matrices,
we have used, just for the ease of writing while writing the
potential, the Pauli matrices themselves.  This adjustment is
inconsequential as far as the potential is concerned: it is just a
matter of redefinition of the concerned coupling constant.

Note that, in this notation,
\begin{eqnarray}
T_\pm^\dagger = - T_\mp \,,  \qquad T_0^\dagger = T_0 \,,
\label{sb.Tdag}
\end{eqnarray}
and
\begin{eqnarray}
\Big[ T_0, T_\pm \Big] = \mp T_\pm \,,
\end{eqnarray}
where $T$ stands for the generators in any representation.  Also, note
that the $T_\pm$ generators differ from the usual definition of the
ladder operators in quantum mechanics texts.

The advantage of working with the spherical components should be
obvious.  The components correspond to fields of definite electric
charge in the $\rm SU(2) \times U(1)$ model.  To see how it helps the
notation, let us look at the term with the coefficient $M_1$
in \Eqn{gm.pot}, and examine one term whose trace appears there:
\begin{eqnarray}
\tr (\Phi^\dagger \tau_-^\dagger \Phi \tau_-) =
\tr \left[  \begin{pmatrix}
    \phi_0 & - \phi_+ \\
    \phi_- & \phi_0^* 
  \end{pmatrix} 
  \begin{pmatrix}
  0 & 0 \\ \surd2 & 0 
\end{pmatrix} 
  \begin{pmatrix}
    \phi_0^* & \phi_+ \\
    - \phi_- & \phi_0 
  \end{pmatrix}
  \begin{pmatrix}
  0 & \surd2 \\ 0 & 0 
\end{pmatrix} 
  \right]  = 2 \phi_0^* \phi_0^* \,.
\end{eqnarray}
Considering the U(1) charge of this combination, it is clear that it
can couple only to $\chi_0$, which is exactly the component $X_{(--)}$
of the matrix $X$, because the rows and columns are supposed to be
marked by the eigenvalues $+1,0,-1$ of $t_0$, in that order.
Similarly, other components can be checked, in the same term as well
as in other terms.  This is much more intuitive than the notation used
in the earlier literature which involves the Cartesian generators, and
therefore the matrix $U$, given in \Eqn{sb.U}, appears explicitly in
the expression for the potential.

Any SU(2) representation is self-conjugate.  It means that, for any
multiplet $\Psi$ with $N$ components, there is a matrix $C_N$ such
that $C_N\Psi^*$ transforms exactly the same way as $\Psi$.  The
matrix $C$ should satisfy the relation
\begin{eqnarray}
CT_\alpha^* = - T_\alpha C
\end{eqnarray}
on the hermitian generators $T_x$, $T_y$ and $T_z$.  In this article,
only 2, 3 and 5 dimensional representations of SU(2) occur.  For these
representations, the conjugation matrices are
\begin{eqnarray}
C_2 = \begin{pmatrix}
0 & 1 \\ -1 & 0
\end{pmatrix} \,, \qquad
C_3 = \begin{pmatrix}
0 & 0 & -1 \\
0 & 1 & 0 \\
-1 & 0 & 0 \\
\end{pmatrix} \,, \qquad
C_5 = \begin{pmatrix}
0 & 0 & 0 & 0 & 1 \\ 
0 & 0 & 0 & -1 & 0 \\
0 & 0 & 1 & 0 & 0 \\
0 & -1 & 0 & 0 & 0 \\
1 & 0 & 0 & 0 & 0 \\
\end{pmatrix} \,.
\label{sb.C23}
\end{eqnarray}
For the doublet defined in \Eqn{sm.phi}, the conjugate doublet has a
minus sign in one of the components, which explains the minus sign
appearing in the first column of the matrix $\Phi$ in \Eqn{gm.PhiX}.
The conjugate triplet of $\chi$ has been defined as
\begin{eqnarray}
\tilde\chi = - C_3 \chi^* \,,
\end{eqnarray}
whose components are shown in \Eqn{gm.chitil}.  The minus sign in
front of $C_3$ is purely a matter of convention.  This $\tilde\chi$
appears in the leftmost column of the matrix $X$ in \Eqn{gm.fields}.

Let us now look at real multiplets.  An arbitary real triplet
$\Sigma$ should satisfy the condition
\begin{eqnarray}
\Sigma = C_3 \Sigma^* \,.
\label{sb.real}
\end{eqnarray}
Equivalently, we can say 
\begin{eqnarray}
\Sigma^\dagger = \Sigma^\top C_3^\dagger = \Sigma^\top C_3 \,.
\end{eqnarray}
Therefore, the scalar product of two real vectors $\Sigma$ and $\Xi$
can be wriiten as 
\begin{eqnarray}
\Sigma^\dagger \Xi = \Sigma^\top C_3 \Xi \,
\end{eqnarray}
which exactly agrees with the expression given in \Eqn{sb.dotprod}.

The condition of \Eqn{sb.real} implies that the middle component of a
real multiplet must be real.  For the other two components, there are
two ways that this condition can be realized.  One is to take
\begin{eqnarray}
\Sigma_{(-)} = - \Sigma_{(+)}^* \,,
\label{sb.real1}
\end{eqnarray}
as suggested by \Eqn{sb.+0-}.  The other possibility is to define a
$\Sigma_+$ (whose subscript is not parenthesized) by the relations
\begin{eqnarray}
 \Sigma_{(+)} = i  \Sigma_{+} \,, \qquad  \Sigma_{(-)} = i  \Sigma_{+}^* \,.
\label{sb.real2}
\end{eqnarray}
In either case, we will write $\Sigma_{+}^*$ as $\Sigma_{-}$, so that
the first kind of triplet will be written with components
$(\Sigma_+, \Sigma_0,-\Sigma_-)$, and the second kind with
$(i\Sigma_+,\Sigma_0, i\Sigma_-)$, where the unparenthesized indices
indicate the electric charge of the concerned field.  For the
5-dimensional representation of the custodial SU(2) of the GM model,
the components are $(F_{++}, iF_+, F_0, iF_-, F_{--})$ where $F_0$ is
real, $F_-=F_+^*$ and $F_{--}=F_{++}^*$.

For various CS multiplets appearing in our discussion, we have
used \Eqn{sb.real1}-type notation for the gauge boson triplet, {\em
i.e.}, we have denoted the CS triplet with components $(-W_+,W_0,W_-)$
with $W_+^*=W_-$.  A similar notation has been used for the
$\xi$-triplet in \Eqn{gm.fields}.  On the other hand, for the
unphysical Goldstone bosons, which also form a CS triplet, we use the
reality condition as in \Eqn{sb.real2}, {\em i.e.}, we have denoted
the components as $(ih_+,h_0,ih_-)$, with $h_+^*=h_-$.
 

\section{Correspondence between parameters in GM, eGM and
the full potential}\label{a:cor}
The full gauge-invariant potential of the model with a doublet, a real
triplet and a complex triplet of Higgs bosons was given in
\Eqn{gm.V16}.  Our notation differs considerably from the notation
used by earlier authors \cite{Keeshan:2018ypw, Das:2018vkv,
  Banerjee:2019gmr}.  The motivation for the change of notation was
explained in the text: our use of two different kinds of quartic
couplings, with two different letters, helps understand the symmetries
of the mass matrices.  In order to bridge between our notation and
previous notation, we do two things here.  First, we provide a
translation chart between different notations in Table~\ref{t:cor}.
Second, we give the full potential in component form, fully expanding
the terms of \Eqn{gm.V16}:
\begin{table}
$$
  \begin{array}{cccc}
    \hline
    \multicolumn2c{\mbox{Full potential}} & \mbox{GM} & \mbox{eGM} \\
    \cline{1-2}
    \mbox{This paper} & \mbox{Ref.\
      \cite{Keeshan:2018ypw}} & \mbox{as in \Eqn{gm.pot}} &
    \\  \hline
    m_\phi^2 & -\tilde{\mu}_2^2 & -m_2^2 &    m_\phi^2 \\
    m_\chi^2 & -\tilde\mu_3^{\prime 2} & -m_3^2 &    m_\chi^2 \\
    m_\xi^2 & -\textstyle{\frac12} \tilde\mu_3^2 & -\frac12 m_3^2 &
    \frac12   m_\chi^2 \\ 
    \mu_1 & -6\tilde M_2 &  -6M_2  &    \mu_1 \\
    \mu_2 & \frac12 \tilde M_1 & 2M_1 &  \surd2 \mu_3 \\
    \mu_3 & \frac1{2\surd2} \tilde M_1^\prime & \surd2 M_1   &    \mu_3 \\
    \lambda_\phi & \tilde{\lambda}_1 & 4 \lambda_1 &    \lambda_\phi \\
    \lambda_\chi & \tilde{\lambda}_7 & 4 \lambda_2 + 2 \lambda_3  &    \lambda_\chi \\
    \lambda_\xi & \tilde{\lambda}_8 & \lambda_2 + \lambda_3 &    \lambda_\xi \\
    \tilde{\lambda}_\chi & \tilde{\lambda}_2 & 2\lambda_3 &    \tilde{\lambda}_\chi \\
    \lambda_{\phi\chi} & \tilde{\lambda}_5 & 4 \lambda_4  &    \lambda_{\phi\chi} \\
    \lambda_{\phi\xi} & \tilde{\lambda}_6 & 2 \lambda_4 &    \lambda_{\phi\xi} \\
    \lambda_{\chi\xi} & \tilde{\lambda}_{10} & 4 \lambda_2  &  2
    \lambda_\chi - 4 \lambda_\xi \\ 
    \kappa_1 & \tilde{\lambda}_9 & 4 \lambda_3 &    \kappa_1 \\
    \kappa_2 & -\textstyle{\frac12} \tilde{\lambda}_3 & 4\lambda_5 &
    \lambda_{\phi\chi} - 2 \lambda_{\phi\xi} - \surd2 \kappa_3 \\ 
    \kappa_3 & \textstyle{\frac12} \tilde{\lambda}_4 & - 2\surd2 \lambda_5 &    \kappa_3 \\
    \hline
\end{array}
  $$
 \caption{Correspondence between different notations.  Columns 1 and
    2 show the notations of the 16-paramter potential.  Column 3 shows
  the parametrization of the GM model potential as given in
  \Eqn{gm.pot} of this paper.  Column 4 shows the restrictions that
  lead to the eGM model proposed in this paper.}\label{t:cor}
  \end{table}
  \allowdisplaybreaks
\begin{eqnarray}
  V(\phi,\chi,\xi) &=&
  - m_\phi^2 (\phi_+\phi_- + \phi_0^*\phi_0)
  - m_\chi^2 (\chi_{++} \chi_{--} + \chi_+ \chi_- + \chi_0^* \chi_0)
  - m_\xi^2 (2\xi_+ \xi_- + \xi_0^2)
  \nonumber \\
  && - \mu_1 \Big[ - (\chi_0 \chi_- + \chi_+ \chi_{--}) \xi_+ +
    (\chi_{++} \chi_{--} - \chi_0^* \chi_0) \xi_0 - (\chi_+ \chi^*_0 +
    \chi_{++}\chi_-) \xi_- \Big] \nonumber \\
  && - \mu_2 \Big[ \surd2 \phi_- \phi_0 \xi_+ + ( \phi_0^*\phi_0 -
    \phi_+ \phi_-) \xi_0 + \surd2 \phi_+ \phi_0^* \xi_- \Big]
  \nonumber\\
  && - \mu_3 \Big[ (\surd2 \phi_0\phi_0\chi^*_0 + 2 \phi_+\phi_0\chi_- +
    \surd2 \phi_+\phi_+\chi_{--}) + \mbox{h.c.} \Big]
  \nonumber\\
  &&+\lambda_\phi \left( \phi_+\phi_- + \phi_0^*\phi_0 \right)^2 +
  \lambda_\chi \left( \chi_{++} \chi_{--} + \chi_+ \chi_- + \chi_0^*
  \chi_0 \right)^2 
  \nonumber\\
  && +   
  \lambda_\xi \left( 2\xi_+ \xi_- + \xi_0^2 \right)^2 +
  \tilde\lambda_\chi \left\vert 2\chi^*_0\chi_{--} - \chi_-\chi_-
  \right\vert^2
  \nonumber\\
  && +\lambda_{\phi\chi} \left( \phi_+\phi_- + \phi_0^*\phi_0 \right)
  \left( \chi_{++} \chi_{--} + \chi_+ \chi_- + \chi_0^* \chi_0 \right)
  \nonumber\\
  && +  \lambda_{\phi \xi} \left( \phi_+\phi_- + \phi_0^*\phi_0 \right)
  \left( 2\xi_+ \xi_- + \xi_0^2 \right)
  + \lambda_{\chi\xi} \left( \chi_{++} \chi_{--} + \chi_+ \chi_- +
  \chi_0^* \chi_0 \right) \left( 2\xi_+ \xi_- + \xi_0^2 \right)
  \nonumber\\
  && + \kappa_1 \left\vert \xi_-\chi^*_0 - \xi_0\chi_- - \xi_+ \chi_{--}
  \right\vert^2
  \nonumber\\
  && + \kappa_2 \Big[ - \surd2 \phi_- \phi_0 (\chi_+\chi^*_0 + \chi_{++}
    \chi_-) + (\phi_+\phi_- - \phi_0^*\phi_0) (\chi_0^*\chi_0 -
    \chi_{++}\chi_{--})
    \nonumber\\*
    && 
     \hspace{4cm}  - \surd2 \phi_+\phi_0^* (\chi_0 \chi_- +
    \chi_+ \chi_{--}) \Big]
  \nonumber\\
  && + \kappa_3 \Big[ \Big\{  \surd2 \phi_0\phi_0 (\xi_0\chi^*_0 + \xi_+\chi_-)
    + 2 \phi_+ \phi_0 (\xi_-\chi^*_0 + \xi_+ \chi_{--}) 
    \nonumber\\*
    && 
     \hspace{4cm} - \surd2
    \phi_+\phi_+ (-\xi_-\chi_- + \xi_0 \chi_{--}) \Big\} + \mbox{h.c.} \Big]\,, 
  \label{cor.V16}
\end{eqnarray}


\section{Renormalization Group Equations}\label{app.rge}
%
%
The RGEs for the trilinear and quartic couplings for the most general
16-parameter model are taken from Ref.\ \cite{Keeshan:2018ypw}, with a
translation to our convention, as shown in Appendix \ref{a:cor}.  The
eGM RGEs are easily obtained by imposing the necessary relationships
among the couplings.  We use the shorthand notation $\beta_A = 16\pi^2
\, dA/dt$ with $t = \ln q$, where $q$ is the relevant energy scale.
Note that the U(1)$_Y$ gauge coupling $g_1$ is not GUT normalized.
\allowdisplaybreaks
\begin{eqnarray}
\beta_{\lambda_\phi} &=& \lambda_\phi \, \left(24\lambda_\phi +
12y_t^2 - \frac95 \, g_1^2-9g_2^2\right) 
-6y_t^4+\frac{27}{200}\, g_1^4+\frac{9}{8}\, g_2^4+\frac{9}{20}\,
g_1^2g_2^2 \nonumber\\* 
&& +2\kappa_2^2+8 \kappa_3^2+3\lambda_{\phi \chi}^2+6\lambda_{\phi \xi}^2\,, \nonumber\\
\beta_{\lambda_\chi} &=& \lambda_\chi \left(16\tilde{\lambda}_\chi + 28\lambda_\chi -\frac{36}{5}\, g_1^2 - 24g_2^2  \right)
+ \frac{54}{25}\, g_1^4 + 9g_2^4 + \frac{36}{5}g_2^2 g_1^2\nonumber\\*
&&  + 16\tilde{\lambda}_\chi^2 +2\lambda_{\phi \chi}^2 + \kappa_1 ^2 + 2 \kappa_2^2 + 2\lambda_{\chi \xi}
\left(3\lambda_{\chi \xi}+2\kappa_1\right)\,, \nonumber\\
\beta_{\tilde{\lambda}_\chi} &=& 
 12\tilde{\lambda}_\chi \left(\tilde{\lambda}_\chi+2\lambda_\chi -\frac{3}{5}\, g_1^2 - 2g_2^2\right) 
+ 3g_2^4 - \frac{36}{5}\, g_1^2 g_2^2  - 2\kappa_2^2+\kappa_1^2\,, \nonumber\\
\beta_{\lambda_\xi} &=& 8\lambda_\xi \left(11\lambda_\xi - 3g_2^2\right) + 3g_2^4 +2\lambda_{\phi \xi}^2
+ \kappa_1 \left( \kappa_1+2\lambda_{\chi \xi}\right) + 3\lambda_{\chi \xi}^2\,, \nonumber\\
\beta_{\lambda_{\phi \chi}}  &=& \lambda_{\phi \chi} \left( 4\lambda_{\phi \chi}+12\lambda_\phi
+8\tilde{\lambda}_\chi+16\lambda_\chi +6 y_t^2 -\frac{9}{2}\, g_1^2 - \frac{33}{2}\, g_2^2 \right)\nonumber\\*
&& +\frac{27}{25}\, g_1^4+6g_2^4+8\kappa_2^2+16 \kappa_3^2+4\lambda_{\phi \xi} \kappa_1
+12 \lambda_{\phi \xi} \lambda_{\chi \xi}\,, \nonumber\\
\beta_{\lambda_{\phi \xi}} &=& 
\lambda_{\phi \xi} \left(8\lambda_{\phi \xi}+12\lambda_\phi+40\lambda_\xi + 6 y_t^2 -\frac{9}{10}\, g_1^2 - 
\frac{33}{2}\, g_2^2\right) + 3g_2^4 + 16\kappa_3^2 + 2\lambda_{\phi \chi} \kappa_1+6\lambda_{\phi \chi}
\lambda_{\chi \xi}\,, 
\nonumber\\
\beta_{\lambda_{\chi \xi}} &=&  2\lambda_{\chi \xi} \left( -\frac{9}{5}\, g_1^2-12g_2^2
+4\tilde{\lambda}_\chi+8\lambda_\chi+20\lambda_\xi+4\lambda_{\chi \xi}\right) + 6g_2^4 \nonumber\\*
&& +8 \kappa_3^2+2\kappa_1^2+
4\lambda_{\phi \chi}\lambda_{\phi \xi}  + 4 \kappa_1 \left(\lambda_\chi+2\lambda_\xi\right)\,,\nonumber\\
\nonumber
\beta_{\kappa_1}  &=& 2\kappa_1 \left(5 \kappa_1+
4\tilde{\lambda}_\chi + 2\lambda_\chi + 8\lambda_\xi + 8\lambda_{\chi \xi}  -\frac{9}{5}\, g_1^2  -12g_2^2 \right) 
+6g_2^4  - 8 \kappa_3^2\,, \\
\nonumber
\beta_{\kappa_2} &=& \kappa_2 \left( 4\lambda_\phi - 8\tilde{\lambda}_\chi + 8\lambda_{\phi \chi} + 4\lambda_\chi
-\frac{9}{2}\, g_1^2 - \frac{33}{2}\, g_2^2 + 6y_t^2 \right)-\frac{18}{5}\, g_1^2g_2^2 - 8\kappa_3^2\,, \\
\nonumber
\beta_{\kappa_3} &=& \kappa_3 \left( 
4\lambda_\phi - 4\kappa_2+4\lambda_{\phi \chi}+8\lambda_{\phi \xi} - 2\kappa_1+4\lambda_{\chi \xi}
+ 6y_t^2 - \frac{27}{10}\, g_1^2 - \frac{33}{2}\, g_2^2 \right)\,, \\
\nonumber
\beta_{\mu_1} &=&\mu_1 \left( - 8\tilde{\lambda}_\chi+4\lambda_\chi - 4\kappa_1
+8\lambda_{\chi \xi}  -\frac{18}{5}\, g_1^2-18g_2^2  \right)+4\mu_2\kappa_2-16 \mu_3 \kappa_3\,, \\
\nonumber
\beta_{\mu_2} &=& \mu_2 \left( 6y_t^2 - \frac{9}{10}\, g_1^2-\frac{21}{2}\, g_2^2+4\lambda_\phi+8\lambda_{\phi \xi}
\right) + 4\mu_1 \kappa_2+32 \mu_3 \kappa_3\,, \\
\beta_{\mu_3} &=& \mu_3(6y_t^2-\frac{27}{10}g_1^2-\frac{21}{2}g_2^2+4\lambda_\phi-8k_2+4\lambda_{\phi \chi})+4k_3(2\mu_2-\mu_1).
\label{RGE}
\end{eqnarray}

\bibliographystyle{unsrt}
\bibliography{bibmerged}

\begin{thebibliography}{10}

\bibitem{ParticleDataGroup:2020ssz}
P.~A. Zyla et~al.
\newblock {Review of Particle Physics}.
\newblock {\em PTEP}, 2020(8):083C01, 2020.

\bibitem{Branco:2011iw}
G.~C. Branco, P.~M. Ferreira, L.~Lavoura, M.~N. Rebelo, Marc Sher, and Joao~P.
  Silva.
\newblock {Theory and phenomenology of two-Higgs-doublet models}.
\newblock {\em Phys. Rept.}, 516:1--102, 2012.

\bibitem{Georgi:1985nv}
Howard Georgi and Marie Machacek.
\newblock {Doubly charged Higgs bosons}.
\newblock {\em Nucl. Phys. B}, 262:463--477, 1985.

\bibitem{Grzadkowski:2010dj}
B.~Grzadkowski, M.~Maniatis, and Jose Wudka.
\newblock {The bilinear formalism and the custodial symmetry in the
  two-Higgs-doublet model}.
\newblock {\em JHEP}, 11:030, 2011.

\bibitem{Haber:2010bw}
Howard~E. Haber and Deva O'Neil.
\newblock {Basis-independent methods for the two-Higgs-doublet model III: The
  CP-conserving limit, custodial symmetry, and the oblique parameters S, T, U}.
\newblock {\em Phys. Rev. D}, 83:055017, 2011.

\bibitem{Solberg:2018aav}
M.~Aa Solberg.
\newblock {Conditions for the custodial symmetry in multi-Higgs-doublet
  models}.
\newblock {\em JHEP}, 05:163, 2018.

\bibitem{Gunion:1990dt}
J.~F. Gunion, R.~Vega, and J.~Wudka.
\newblock {Naturalness problems for $\rho = 1$ and other large one loop effects
  for a standard model Higgs sector containing triplet fields}.
\newblock {\em Phys. Rev. D}, 43:2322--2336, 1991.

\bibitem{Chiang:2012cn}
Cheng-Wei Chiang and Kei Yagyu.
\newblock {Testing the custodial symmetry in the Higgs sector of the
  Georgi-Machacek model}.
\newblock {\em JHEP}, 01:026, 2013.

\bibitem{Blasi:2017xmc}
Simone Blasi, Stefania De~Curtis, and Kei Yagyu.
\newblock {Effects of custodial symmetry breaking in the Georgi-Machacek model
  at high energies}.
\newblock {\em Phys. Rev. D}, 96(1):015001, 2017.

\bibitem{Keeshan:2018ypw}
Ben Keeshan, Heather~E. Logan, and Terry Pilkington.
\newblock {Custodial symmetry violation in the Georgi-Machacek model}.
\newblock {\em Phys. Rev. D}, 102(1):015001, 2020.

\bibitem{Logan:2015xpa}
Heather~E. Logan and Vikram Rentala.
\newblock {All the generalized Georgi-Machacek models}.
\newblock {\em Phys. Rev. D}, 92(7):075011, 2015.

\bibitem{Palbook}
See, e.g., Palash B. Pal, {\em An introductory course of Particle Physics},
  (CRC Press, 2014), Section~19.5.

\bibitem{Kundu:1995qb}
Anirban Kundu and Biswarup Mukhopadhyaya.
\newblock {A General Higgs sector: Constraints and phenomenology}.
\newblock {\em Int. J. Mod. Phys. A}, 11:5221--5244, 1996.

\bibitem{Aoki:2007ah}
Mayumi Aoki and Shinya Kanemura.
\newblock {Unitarity bounds in the Higgs model including triplet fields with
  custodial symmetry}.
\newblock {\em Phys. Rev. D}, 77(9):095009, 2008.
\newblock [Erratum: Phys.Rev.D 89, 059902 (2014)].

\bibitem{Hartling:2014zca}
Katy Hartling, Kunal Kumar, and Heather~E. Logan.
\newblock {The decoupling limit in the Georgi-Machacek model}.
\newblock {\em Phys. Rev. D}, 90(1):015007, 2014.

\bibitem{CMS:2017fgp}
Albert~M Sirunyan et~al.
\newblock {Search for Charged Higgs Bosons Produced via Vector Boson Fusion and
  Decaying into a Pair of $W$ and $Z$ Bosons Using $pp$ Collisions at
  $\sqrt{s}=13\text{ }\text{ }\mathrm{TeV}$}.
\newblock {\em Phys. Rev. Lett.}, 119(14):141802, 2017.

\bibitem{CMS:2021wlt}
Albert~M Sirunyan et~al.
\newblock {Search for charged Higgs bosons produced in vector boson fusion
  processes and decaying into vector boson pairs in proton\textendash{}proton
  collisions at $\sqrt{s} = 13\,{\text {TeV}} $}.
\newblock {\em Eur. Phys. J. C}, 81(8):723, 2021.

\bibitem{Ismail:2020kqz}
Ameen Ismail, Ben Keeshan, Heather~E. Logan, and Yongcheng Wu.
\newblock {Benchmark for LHC searches for low-mass custodial fiveplet scalars
  in the Georgi-Machacek model}.
\newblock {\em Phys. Rev. D}, 103(9):095010, 2021.

\bibitem{Richard:2021edf}
Richard, Fran\c{c}ois, {\em Global interpretation of LHC indications within the
  Georgi-Machacek Higgs model}, Talk presented at the International Workshop on
  Future Linear Colliders (LCWS2021), 15-18 March 2021, arXiv:2103.12639.

\bibitem{Crivellin:2021ubm}
Andreas Crivellin, Yaquan Fang, Oliver Fischer, Abhaya Kumar, Mukesh Kumar,
  Elias Malwa, Bruce Mellado, Ntsoko Rapheeha, Xifeng Ruan, and Qiyu Sha.
\newblock {Accumulating Evidence for the Associate Production of a Neutral
  Scalar with Mass around 151 GeV}.
\newblock arXiv:2109.02650.

\bibitem{Richard:2021ovc}
Richard, Fran\c{c}ois, {\em A Georgi-Machacek Interpretation of the Associate
  Production of a Neutral Scalar with Mass around 151 GeV}, ILC Workshop on
  Potential Experiments, arXiv:2112.07982.

\bibitem{Das:2018vkv}
Dipankar Das and Ipsita Saha.
\newblock {Cornering variants of the Georgi-Machacek model using Higgs
  precision data}.
\newblock {\em Phys. Rev. D}, 98(9):095010, 2018.

\bibitem{Banerjee:2019gmr}
Avik Banerjee, Gautam Bhattacharyya, and Nilanjana Kumar.
\newblock {Impact of Yukawa-like dimension-five operators on the
  Georgi-Machacek model}.
\newblock {\em Phys. Rev. D}, 99(3):035028, 2019.

\end{thebibliography}

\end{document}